\documentstyle[aps,prc,epsfig,multicol]{revtex}
\newcommand{\etap}{\eta^\prime}
\newcommand{\epsi}{\varepsilon}

\def\Id{{\rm 1\kern-.3em I}}

\begin{document} 

\preprint{TK--00--01}
  
\title{\Large\bf\sc A Relativistic Quark Model for Mesons \\ 
        with an Instanton--Induced Interaction} 

\author{ Matthias Koll\thanks{e-mail: {\tt koll@itkp.uni-bonn.de}}, Ralf
  Ricken, Dirk Merten, Bernard Metsch, Herbert
  Petry \\ {\bigskip}  {\sl Institut f\"ur Theoretische Kernphysik \\
  Nu{\ss}allee 14--16, D--53115 Bonn \\ Germany} \\ {\bigskip} }

\date{\today}

\maketitle

\begin{abstract}

We present new results of a relativistic quark model based on the
Bethe--Salpeter equation in its instantaneous approximation. 
Assuming a linearly rising confinement potential with an
appropriate spinorial structure in Dirac space and adopting a residual
interaction based on instanton effects, we can compute
masses of the light mesons up to highest observed angular momenta with
a natural solution of the $U_A(1)$ problem. The calculated ground
states masses and  the radial excitations describe the experimental results well.
In this paper, we will also discuss our results concerning numerous
meson decay properties. For processes like
$\pi^+/K^+\to e^+\nu _e\gamma$ 
and $0^-\to\gamma\gamma$ at various photon virtualities, 
we find a good agreement with experimental data. We will also comment
on the form factors of the $K_{\ell 3}$ decay and on the decay
constants of the $\pi$, $K$ and $\eta$ mesons. For the sake of
completeness, we will furthermore present the electromagnetic form
factors of the charged $\pi$ and $K$ mesons as well as a comparison of the
radiative meson decay widths with the most recent experimental data.

\end{abstract}

\section{Introduction}

After years of research on the problem of bound states in QCD, there
are still lots of open questions. Since it is not possible to apply a
perturbative treatment of QCD in the low--energy region
around $\approx 1$GeV, one has to rely on effective theoretical
descriptions of hadrons. Integrating out the quark degrees of freedom
leads to approaches such as Chiral Perturbation Theory or
the Nambu--Jona-Lasinio model, but these ideas fail in the description
of higher lying resonances and radial excitations. These states can be
described (even in a non--relativistic treatment at least qualitatively) 
by models including quark confinement (see {\it e.g.} \cite{GodfreyIsgur}).

In this paper, we discuss some new results on light meson spectra and decays in
the framework of a relativistic quark model that has been presented in
some previous publications (see \cite{MuenzPetry1} -- \cite{GierscheMuenz}). 
In particular, we want to update our results on mass spectra and
electroweak decay properties, consistently calculated with a parameter
set that gives a global description of the complete meson spectrum. At
the same time, we shall discuss new results from an alternative
description of confinement. Our model 
is based on the Bethe--Salpeter equation in its instantaneous
approximation and provides an excellent description of the
masses of the complete meson spectrum including highest angular
momenta and radial excitations. We will study these spectra
and compare our results not only with the latest {\sc Particle Data Group} (PDG)
compilation \cite{PDG00} but also with alternative new interpretations
of meson resonances as $q\bar q$ states or exotics.  
The model that is presented here can also be applied to
various meson decay processes and shows a good
overall agreement compared to the experimental data. We will
investigate the pseudoscalar decay constants and their relation to the
decays of $J^\pi=0^-$ mesons into two photons at some selected photon
virtualities. The electromagnetic structure of the charged $\pi$ and
$K$ mesons will be discussed by presenting their form factors,
calculated in the parameter sets used in this paper. Since we want to
update our former publications, we will also briefly resume the status
of the electromagnetic decay widths in our model. 
Furthermore, we compute form factors for the processes 
$\pi^+/K^+\to e^+\nu _e\gamma$ and
$K^+\to\pi^0 e^+ \nu _e$ (the so--called $K_{\ell 3}$ decay). 

We have organized this article as follows: Section \ref{BetheSalpeter}
gives a synopsis of our model and introduces the potentials
adopted in the subsequent evaluations. Section
\ref{ParametersSpectra} is devoted to a discussion of the parameters
and their effects on the resulting meson spectra. We will present
calculations on various decay processes in section
\ref{DecayProperties} before we conclude with summarizing remarks in
section \ref{SummaryOutlook}.

\section{A Relativistic Quark Model on the Basis 
      \\ of the Instantaneous Bethe--Salpeter Equation} 
                                        \label{BetheSalpeter}

\begin{figure}[h]
  \protect\begin{center}
    \leavevmode
       \protect\input{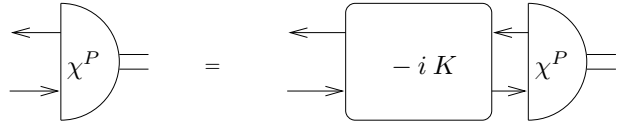}
\newline
       \protect\caption{The Bethe--Salpeter equation for $q\bar q$ bound
    states in a graphical notation.}
       \protect\label{fig:BSEqu}
   \end{center}
\end{figure}

Our results will be presented in the framework of a relativistic quark
model based on the Bethe--Salpeter (BS) equation (see
\cite{SalpeterBethe}) for a $q\bar q$ bound state of four-momentum $P$ and mass
$M$ with $M^2=P^2$:  
\begin{eqnarray}
  \label{BSE}
  \chi^{P} (p) = -i \: S_1^F (\eta _1 P+p) \left[
  \int \frac{d^4 p'}{(2\pi)^4} K(P; p, p') \chi^{P} (p') \right]
  S_2^F(-\eta_2 P+p)  \quad ; 
\end{eqnarray}
see fig. (\ref{fig:BSEqu}) for a diagrammatic representation of this
equation. Here, $S^F_i(\pm \eta_iP+p)$ denotes the full quark
propagator where $i=1$ indicates the quark, $i=2$ the
antiquark and $p$ is the relative momentum between quark and
antiquark.  The coefficients $\eta _i$  satisfying the condition
$\eta _1 + \eta _2 = 1$ fix a special choice 
of coordinates; in the following considerations, we will set $\eta
_1=\eta _2=\frac 1 2$ for the sake of simplicity. 

The BS amplitude $\chi ^P$ is defined in
coordinate space as the time--ordered product of two (anti--)quark
field operators: 
\begin{eqnarray}
      \chi ^{P}_{\alpha\beta} (x_1, x_2) &:=&  \left\langle\: 0 \: \left |
 T \:\psi_\alpha^1 (x_1)\bar\psi^2_\beta (x_2)\right |\:P
 \:\right\rangle\nonumber 
\\ &=& e^{-iP\cdot\frac{x_1 + x_2}{2}} \int\frac{d^4 p}{(2\pi)^4}
 e^{-ip\cdot(x_1 - x_2)} \chi ^{P}_{\alpha\beta} (p) \quad ,
\end{eqnarray}
where $\alpha$ and $\beta$ stand for Dirac, flavour and colour
indices. The function $K(P;p,p')$ in the BS equation 
represents the four--dimensional irreducible kernel
including all interactions between the $q\bar q$ pair. Neither the
full propagator $S_i^F(\pm P/2+p)$ nor the interaction kernel
$K(P;p,p')$ are sufficiently well known and have to be fixed by
appropriate phenomenological assumptions. In our model, we adopt the so--called
instantaneous approximation for the kernel that was originally proposed by Salpeter
(see \cite{Salpeter}). It can be formulated covariantly via 
\begin{equation}
   K(P; p, p') = V (p_{\bot}, p'_{\bot}) \quad ,
\end{equation}
where we have introduced components of the relative momentum $p=p_{\|} +
p_{\bot}$ parallel and perpendicular to the meson momentum $P$ by 
\begin{equation}
p_{\|} := \frac{p\cdot P}{\sqrt{P^2}}\quad\mbox{and}\quad
p_{\bot} := p -  \frac{p\cdot P}{\sqrt{P^2}} \quad .
\end{equation}
In the rest frame of the meson where $P=(M, \vec 0)$, we find
$p_\|=(p^0, \vec 0)$ and $p_\bot = (0,\vec p)$ yielding finally
\begin{equation}
  \left. K(P; p, p') \right | _{P=(M, \vec 0)} = V (\vec p,  \vec p\,') 
\end{equation}
for the instantaneous interaction kernel.  This formally covariant
formulation allows to transform any solution $\chi^P$ of the BS
equation that is found in the  
meson rest frame with $P=(M, \vec 0)$ into a solution for non--vanishing meson
momenta. This is an important point since it turned out to be crucial
for a satisfying description of {\it e.g.} the pion form factor already at
moderate $Q^2$ that indeed the correct Lorentz boost is applied to
the BS amplitudes of the $\pi$ meson (see \cite{MuenzPetry3}).

The second model assumption states that the quark propagators in the
BS equation can suitably be approximated by free propagators
according to
\begin{eqnarray}
  \label{QuarkPropagator}
  S^F_i (p) &\approx& i \:\frac{\not p + m_i}{p^2 - m_i^2} 
             =    i \:\left(\frac{\Lambda _i^+ (\vec p)}{p^0 - \omega _i +
          i\epsi}  + \frac{\Lambda _i^- (\vec p)}{p^0 + \omega _i - i\epsi }
          \right)\gamma ^0 
\end{eqnarray}
where $m_i$ is the effective constituent mass 
(either for {\sl nonstrange} or {\sl strange} flavours in the case
of light mesons) with $\omega _i = \sqrt{\vec p\, ^2 +
  m_i^2}$ the energy of the (anti--)quark $i$. The projectors
\begin{equation}
  \Lambda _i^\pm (\vec p) := \frac 1 2 \pm \frac{H_i (\vec p)}{2\omega _i}
\end{equation}
with the standard single particle Dirac Hamiltonian $H_i(\vec p) = \gamma ^0 \left(
  \vec\gamma\vec p + m_i\right)$ distinguish states of
positive and negative energy of (anti--)quark $i$.  

With these assumptions and since the $p^0$ dependence of the interaction
kernel vanishes in the instantaneous approximation, the $p^0$
integration of eq. (\ref{BSE}) in the meson's rest frame can be
performed analytically via the
residue theorem thus leading to the so--called Salpeter equation (see\cite{Salpeter}): 
\begin{eqnarray}
 \Phi (\vec p) &=& + \:\Lambda_1^-(\vec p) \gamma ^0 \left[ \int\frac{d^3
 p'}{(2\pi)^3} \frac{V(\vec p, \vec p\, ') \Phi (\vec p\, ')}{M+\omega _1 +
 \omega _2} \right] \gamma ^0\Lambda _2^+
 (-\vec p) \nonumber \\
 & & -\: \Lambda_1^+(\vec p) \gamma ^0 \left[ \int\frac{d^3
 p'}{(2\pi)^3}\frac{ V(\vec p, \vec p\, ') \Phi (\vec p\, ')}{M-\omega _1 -
 \omega _2} \right] \gamma ^0\Lambda _2^-  (-\vec p) \quad .
\end{eqnarray}
Here, the Salpeter amplitude defined by $\Phi (\vec p) := \int\frac{dp^0}{2\pi}
\:\left.\chi ^P (p^0, \vec p)\right|  _{P=(M, \vec 0)}$ depends only on
the space--like components of the meson's relative momentum $p$.

In previous papers, it has been shown how to formulate the Salpeter
equation as an eigenvalue problem
\begin{equation}
\label{EigenvalueProblem}
({\cal H} \Psi)(\vec p) = M \Psi(\vec p)
\end{equation}
with $\Psi(\vec p) := \Phi (\vec p) \gamma ^0$ and $M$ the mass
of the the $q \bar q$ bound state considered. For
more details, we refer the reader to \cite{MuenzPetry1} and \cite{MuenzPetry2} where also
the numerical treatment of this eigenvalue equation is discussed.

With an adequate potential ansatz, it is thus possible to
calculate meson mass spectra on the basis of the Salpeter
equation. Since we wish a proper description not only of the 
Regge trajectories $M^2\propto J$ but also of the intriguing scalar sector and the
characteristic pseudoscalar splittings, we do not consider potentials derived from
(flavour independent) one--gluon--exchange diagrams but we adopt the
following to describe the underlying quark dynamics: 
\begin{itemize}
\item A linear confinement potential with ${\cal V}_C (x) = a_c + b_c\cdot
  	x$ in coordinate space and an appropriate spinorial structure
	$\mit\Gamma\otimes\mit\Gamma$ in Dirac space acting like
	   \begin{equation}
	    \int\frac{d^3 p'}{(2\pi)^3}  V_C (\vec p, \vec p\, ')\Phi(\vec p\, ') 
	=  \int\frac{d^3 p'}{(2\pi)^3} \tilde{\cal V}_C \left(\left(\vec
	       p - \vec p\, '\right)^2 \right) {\mit\Gamma} \Phi (\vec
	p\, ') {\mit\Gamma} 
	   \end{equation}
	in momentum space. The confinement offset $a_c$ and its slope $b_c$ are free
	parameters of our model. Various spin dependencies have been
	investigated. Below, we will discuss two variants that both yield a
	stable solution of the Salpeter equation and at the same time
	reproduce the states on the Regge trajectories correctly.	
\item A flavour dependent two--body force from an instanton induced
	interaction (abbr.: {\sc iii}),
	following an idea of 't Hooft (see \cite{tHooft},
	\cite{MuenzPetry2} and references therein):
 	  \begin{eqnarray}
	\lefteqn{\int\frac{d^3 p'}{(2\pi)^3}  V_{\mbox{\scriptsize III}}
	    (\vec p, \vec p\, ')\Phi(\vec p\, ')} \\
	&=& 4 G(g,g') \int\frac{d^3 p'}{(2\pi)^3} {\cal R}_\Lambda \left(\vec
 	      p, \vec p\, ' \right) \Big(\Id \mbox{tr}\big[\Phi (\vec p\,
	    ')\big] +  \gamma_5 \mbox{tr}\big[\Phi (\vec p\, ')\gamma
 	   _5\big]\Big) \nonumber\quad .
 	  \end{eqnarray}
	 Here,  ${\cal R}_\Lambda$ represents a regularizing function and
 	 $G(g,g')$ is a flavour matrix, {\it i.e.} a summation over flavour indices
	is understood. We treat the coupling strengths $g$
 	      ({\sl nonstrange} sector), $g'$ ({\sl nonstrange}/{\sl strange} sector) and the 
  	finite effective range $\Lambda=\Lambda _{\mbox{\scriptsize III}}$ as free parameters.
\end{itemize}
The latter feature enables us to describe properly the $\pi$--$K$--$\eta$--$\etap$
mass splittings; without an explicit flavour dependent residual interaction,
their masses would be partly degenerate.

Let us finally introduce the meson--quark--antiquark vertex function
$\Gamma ^P$ (or the amputated BS amplitude) which is defined by 
\begin{equation}
\label{VertexFunction}
  \Gamma ^P (p) := \left[ S_1^F(\frac P 2 + p)\right]^{-1} \:\chi^P(p)\: 
                   \left[ S_2^F(-\frac P 2 + p)\right]^{-1}   \quad .
\end{equation}
It depends only on variables $p_\bot$ of a three--dimensional subspace
and therefore reduces in the rest frame of the meson to
\begin{equation}
\label{VertexFunction2}
  \left. \Gamma ^P(p)\right| _{P=(M, \vec 0)} 
  = -i\int\frac{d^3 p'}{(2\pi)^3} V(\vec p, \vec  p\, ') \Phi(\vec p\, ')
  =: \Gamma (\vec p\:)
\end{equation}
which can be seen by inserting $\Gamma^P (p)$ in the
BS equation with an instantaneous interaction kernel. After simultaneously
computing the mass spectra and the associated Salpeter amplitudes with
eq. (\ref{EigenvalueProblem}), it is therefore possible to reconstruct
the BS amplitudes $\chi ^P$ in the meson's rest frame with
eqs. (\ref{VertexFunction}) and (\ref{VertexFunction2}).

It has been shown in \cite{MuenzPetry3} that --- for a pure boost defined
by $P=\Lambda _P \tilde P$  with $\tilde P=(M,\vec 0)$ --- the rest frame
BS amplitude $\chi ^{\tilde P}$  is linked via $\chi ^P (p) = S_{\Lambda _P}\: \chi ^{\tilde P} (\Lambda _P^{-1} \: p)\: S_{\Lambda _P}^{-1}  $
to the amplitude $\chi ^P$ for any on--shell momentum $P$ ($P^2=M^2$) of the
$q\bar q$ bound state considered; here, $S_{\Lambda _P}$ is a matrix
acting on the spinor indices of $\chi ^P$ that obeys
$S_{\Lambda _P}\gamma _\mu S_{\Lambda _P}^{-1} = (\Lambda _P)_\mu
{}^\nu\gamma_\nu$.

A similar relation holds for the vertex function $\Gamma ^P$ so that
we regard our model as fully relativistic (and not only 'relativized')
as it fulfills the general prescriptions for Lorentz boosts. This can
be seen as a consequence of the covariant formulation of our ansatz
which is in fact possible in spite of the use of the instantaneous
kernels in the BS equation as has been shown above.

\section{Parameters and Mass Spectra}
                                        \label{ParametersSpectra} 

As discussed in the previous section, the relativistic quark
model presented here contains some free parameters:
the effective constituent quark masses $m_n$ and $m_s$, the
confinement parameters $a_c$ and $b_c$ (together with an appropriate
spin structure), and the couplings $g$ and $g'$ for 't Hooft's
instanton--induced force with an effective range $\Lambda
_{\mbox{\scriptsize III}}$. We want to stress that the latter residual
interaction only acts on mesons with $J=0$ as has been stated in
\cite{MuenzPetry2}. Accordingly we can apply the following scheme for
parameter fixing: 
\begin{enumerate}
\item Choose the quark masses in a physically reasonable range,
{\it i.e.} $m_n\approx 300\ldots 400$MeV and $m_s\approx 500\ldots
600$MeV. 
\item Assume an appropriate spin structure that does not
conflict with the condition of numerical stability and that provides the
correct position of states on the Regge trajectories. 
\item Fit the confinement offset and slope to the mesons with
$J\not=0$ and do some fine tuning of the constituent quark masses.  
\item Observe the mass spectrum in the scalar and pseudoscalar sector
for $g,g'\not=0$ and fix these coupling constants to the
$\pi$--$K$--$\eta$--$\eta '$ mass splittings.
\end{enumerate}
It is crucial for the description of the light meson sector with $J=0$
that a flavour dependent force lifts the degeneracies which would
otherwise occur in models only assuming confinement and/or
one--gluon--exchange potentials. Therefore we include 't Hooft's
instanton induced interaction in the parameter sets of our two model
variants, see table \ref{tab:Parameters}. They mainly differ in the 
form assumed for the spin dependence of the confinement force: in
model $\cal A$, a combination of a scalar and a timelike vector
structure is adopted as in ref. \cite{KlemptMuenz} whereas model $\cal
B$ employs a Fierz invariant and $\gamma _5$ invariant spin dependence
also investigated by B\"ohm {\it et al.} (see \cite{BoehmJoosKrammer}) as
well as by Gross and Milana  (see \cite{GrossMilana}). A more detailed
discussion (especially focussing the radial excitations) will be
presented in a separate contribution \cite{Ricken1}. 

\begin{figure}
  \protect\begin{center}
    \leavevmode
       \protect\input{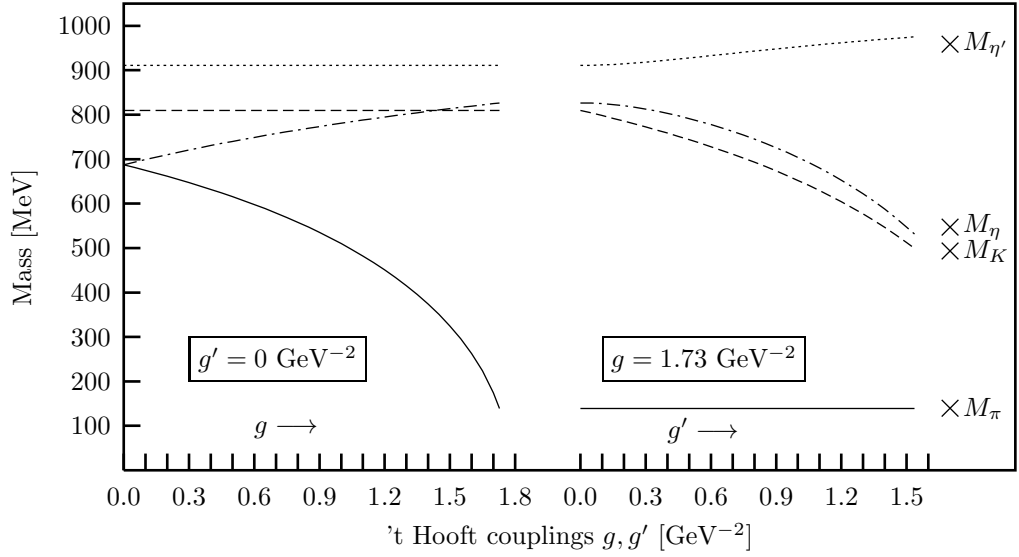}
\newline
       \protect\caption{The effect of 't Hooft's instanton induced
  interaction on the masses of the pseudoscalar mesons with the
  parameters of model $\cal A$. Solid line:
  $M_\pi$; dashed line: $M_K$; dashed--dotted line: $M_\eta$; dotted
  line: $M_{\eta'}$; crosses denote the experimental masses from
  the {\sc Particle Data Group} (see \protect\cite{PDG00}).}
      \protect \label{fig:tHooftEffectA}
   \end{center}
\end{figure}

\begin{figure}
  \protect\begin{center}
    \leavevmode
       \protect\input{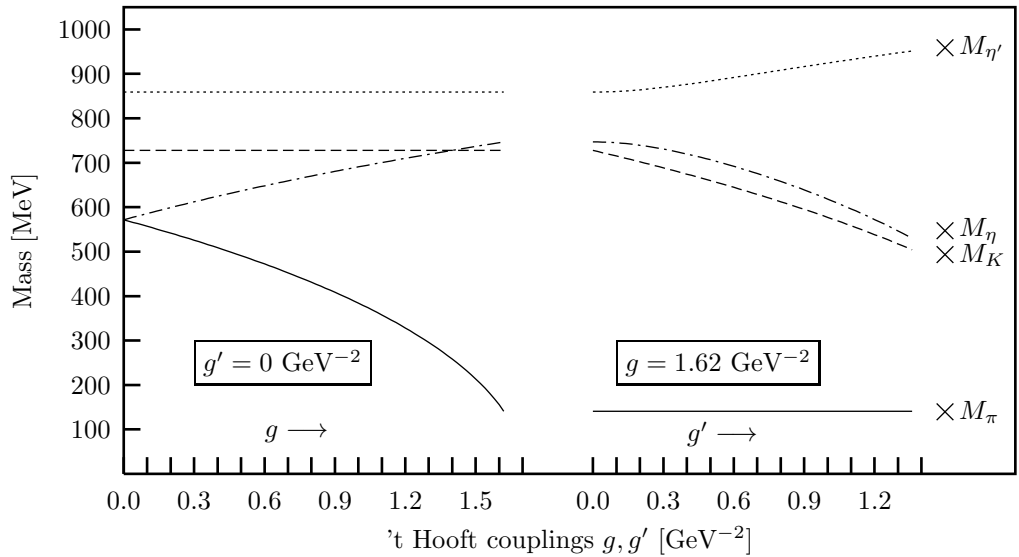}
\newline
       \protect\caption{The effect of 't Hooft's instanton induced
  interaction on the masses of the pseudoscalar mesons with the
  parameters of model $\cal B$. See also caption of fig. (\ref{fig:tHooftEffectA}).}
      \protect \label{fig:tHooftEffectB}
   \end{center}
\end{figure}

In figs. (\ref{fig:tHooftEffectA}) and (\ref{fig:tHooftEffectB}), the
effect of the instanton induced interaction $V_{\mbox{\scriptsize III}}$ is
shown for the light pseudoscalar mesons with the parameters of both
models. For $g=0$ and $g'=0$, the pseudoscalar
mesons are bound by confinement only. Therefore the $\pi$ and the
$\eta$ are degenerate in this limit since no mixing is induced in the isoscalar
sector; in a calculation without 't Hooft's interaction, the $\eta$ is
therefore a pure {\sl nonstrange} state ({\it i.e.} $|n\bar n\rangle = (|u\bar u\rangle
+ |d\bar d\rangle)/\sqrt 2$). With 
increasing coupling $g$, the pion mass is lowered to its experimental value
while $M_\eta$ grows since the underlying effective interaction
has an opposite sign for isovector and isoscalar states (see
\cite{MuenzPetry2} for details). With fixed $g$, the {\sl
nonstrange}--{\sl strange}
coupling $g'$ is used to bring simultaneously the masses of the $K$,
$\eta$ and $\eta '$ mesons to their experimental values. We stress that
a non--vanishing coupling $g'$ not only yields the correct
$\eta$--$\eta '$ mass splitting but also induces the expected
{\sl nonstrange}--{\sl strange} mixing in the isoscalar sector, see
also table \ref{tab:XYCoefficients} for explicit results. 

\begin{figure}
  \protect\begin{center}
    \leavevmode
       \protect\input{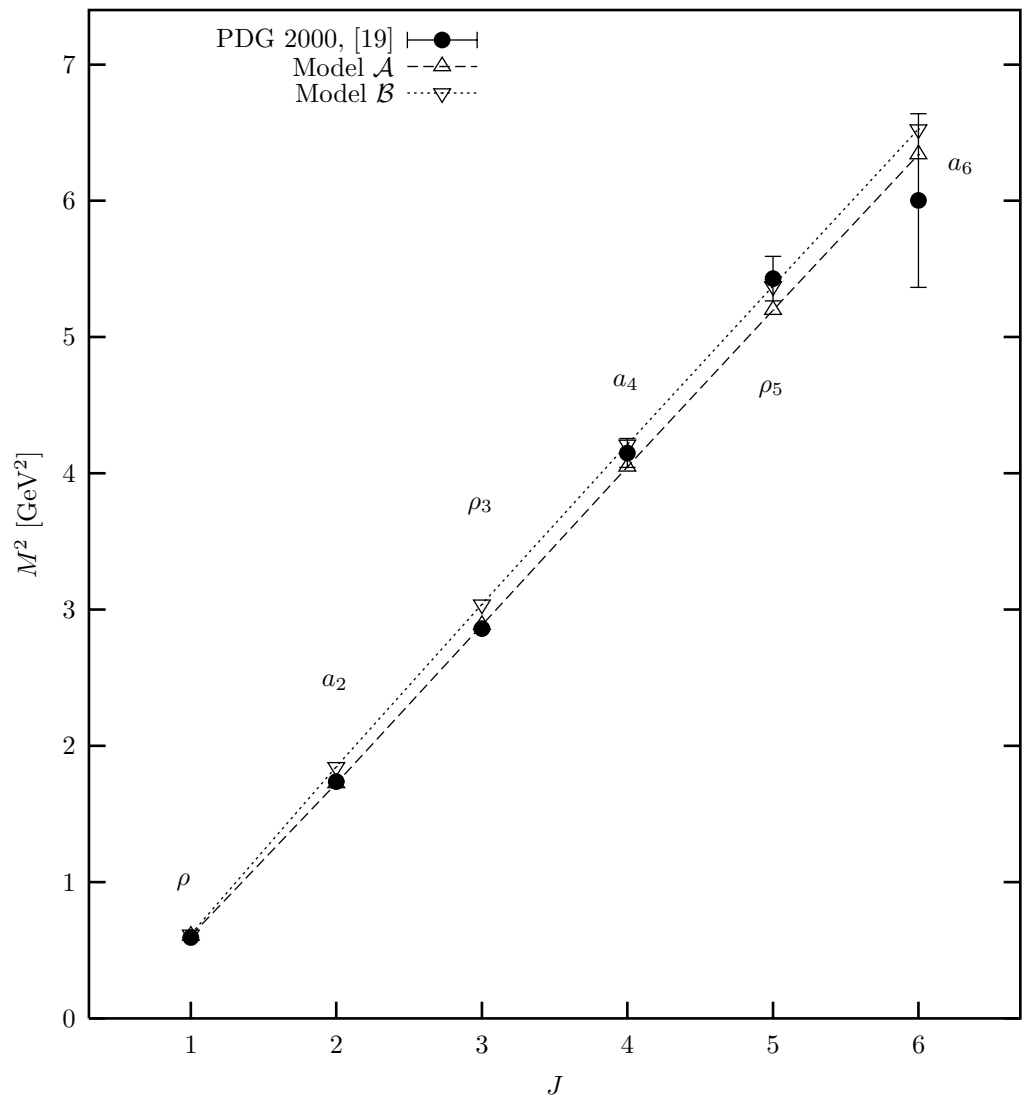}
\newline
       \protect\caption{The Regge trajectory for light isovector mesons with
  the parameters of model $\cal A$ and model $\cal B$ compared to
  experimental masses from the {\sc Particle Data Group} (see \protect\cite{PDG00}).}
      \protect \label{fig:Reggeplot}
   \end{center}
\end{figure}

\begin{figure}
  \protect\begin{center}
    \leavevmode
       \protect\input{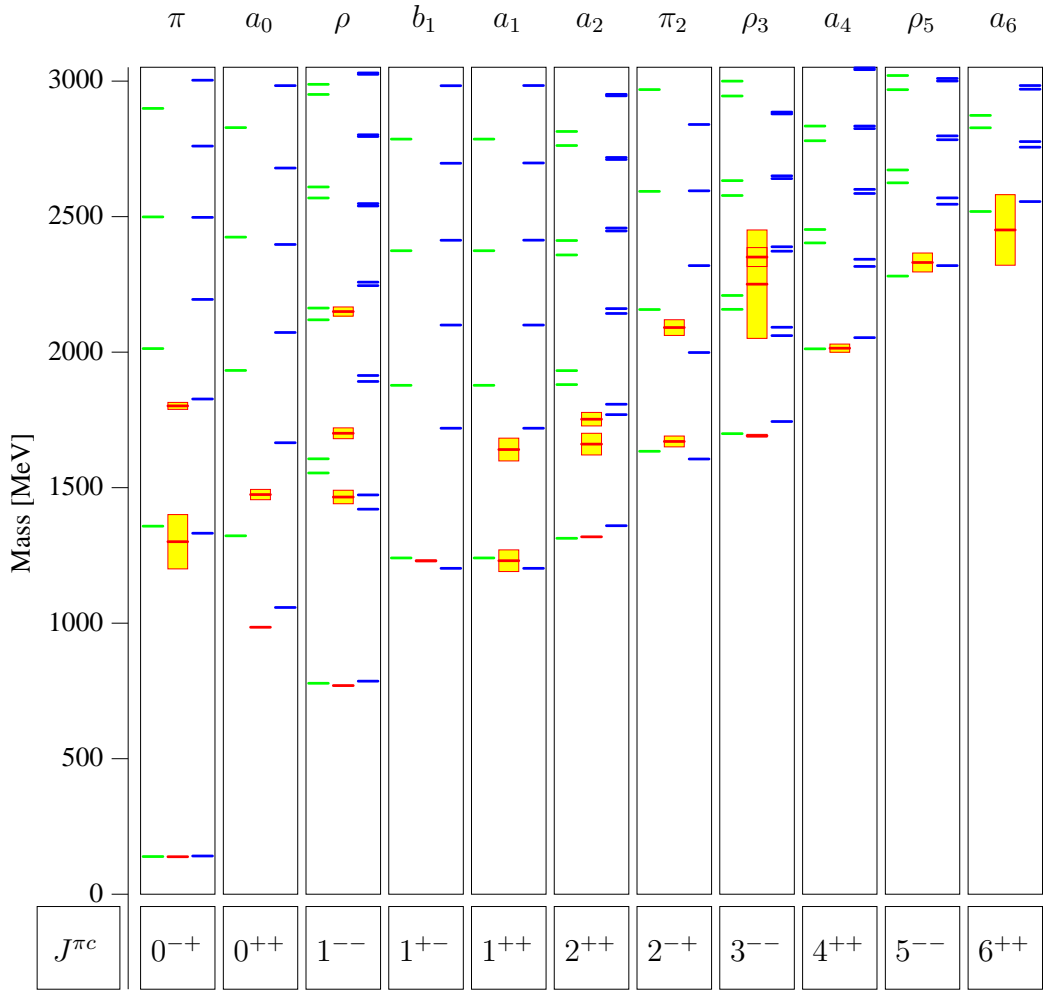}
\newline
       \protect\caption{The spectrum of the light mesons with isospin $I=1$. Left
    column for each $J^{\pi c}$: model $\cal A$; middle column for
    each $J^{\pi c}$: experimental masses and their error bars marked
    by the shadowed rectangles from 
  the {\sc Particle Data Group} (see \protect\cite{PDG00}); right
  column for each $J^{\pi c}$: model $\cal B$. Note the difference for the $J^{\pi c}=0^{++}$ states in the two parameter sets.}
      \protect \label{fig:IsoVector}
   \end{center}
\end{figure}

\begin{figure}
  \protect\begin{center}
    \leavevmode
       \protect\input{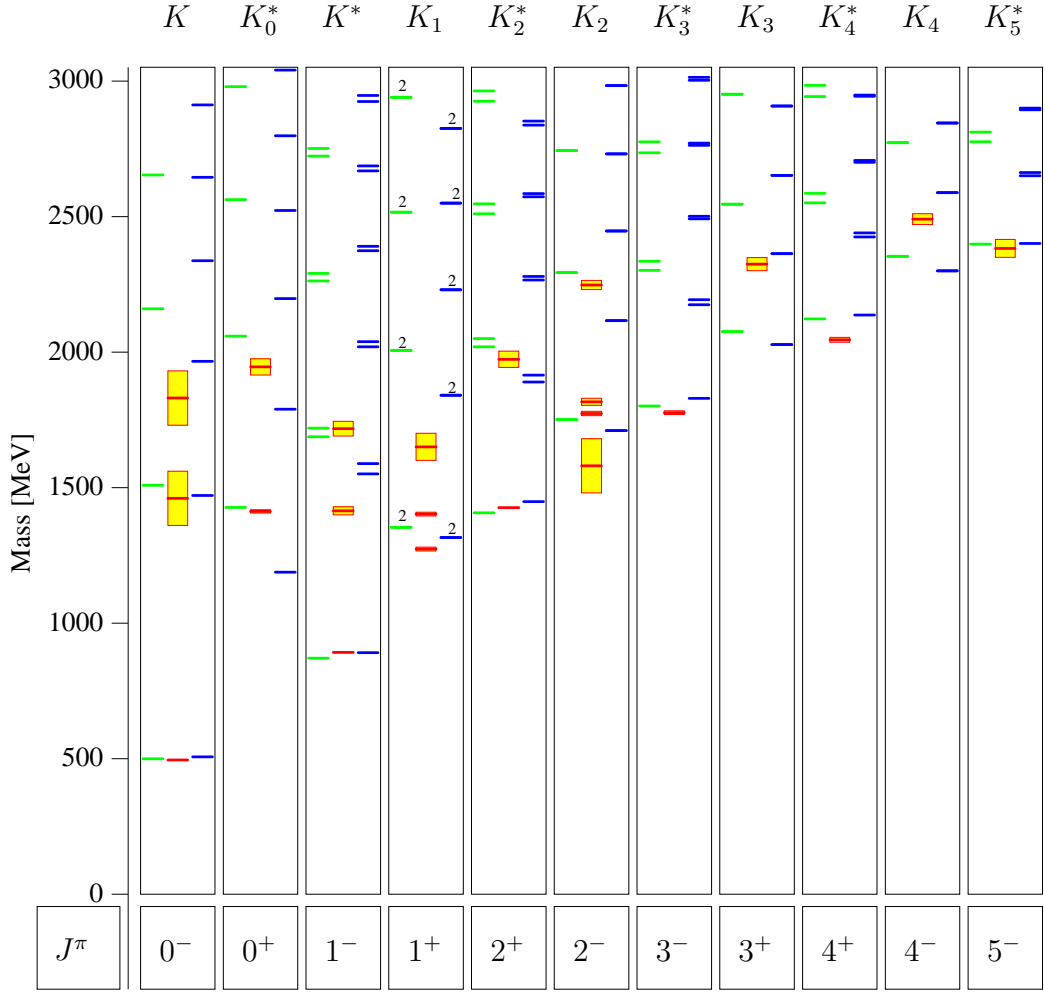}
\newline
       \protect\caption{The spectrum of the light mesons with isospin
    $I=\frac 1 2$. Left
    column for each $J^{\pi}$: model $\cal A$; middle column for
    each $J^{\pi}$: experimental masses and their error bars marked
    by the shadowed rectangles from 
  the {\sc Particle Data Group} (see \protect\cite{PDG00}); right
  column for each $J^{\pi}$: model $\cal B$. The experimental
  data for the $K_3$ and $K_4$ masses need confirmation; the PDG data
  plotted above do not fit in a linear Regge trajectory $M^2_{K_J}\propto
    J$. Note that the calculated $K_1$ states are each 2--fold degenerate for
    spin $S=0$ and $S=1$, indicated by ``2'', so that the total number of $K_1$ states is
    correct; see the discussion in section \protect\ref{ParametersSpectra}.}
       \protect\label{fig:IsoDublett}
   \end{center}
\end{figure}

\begin{figure}
  \protect\begin{center}
    \leavevmode
       \protect\input{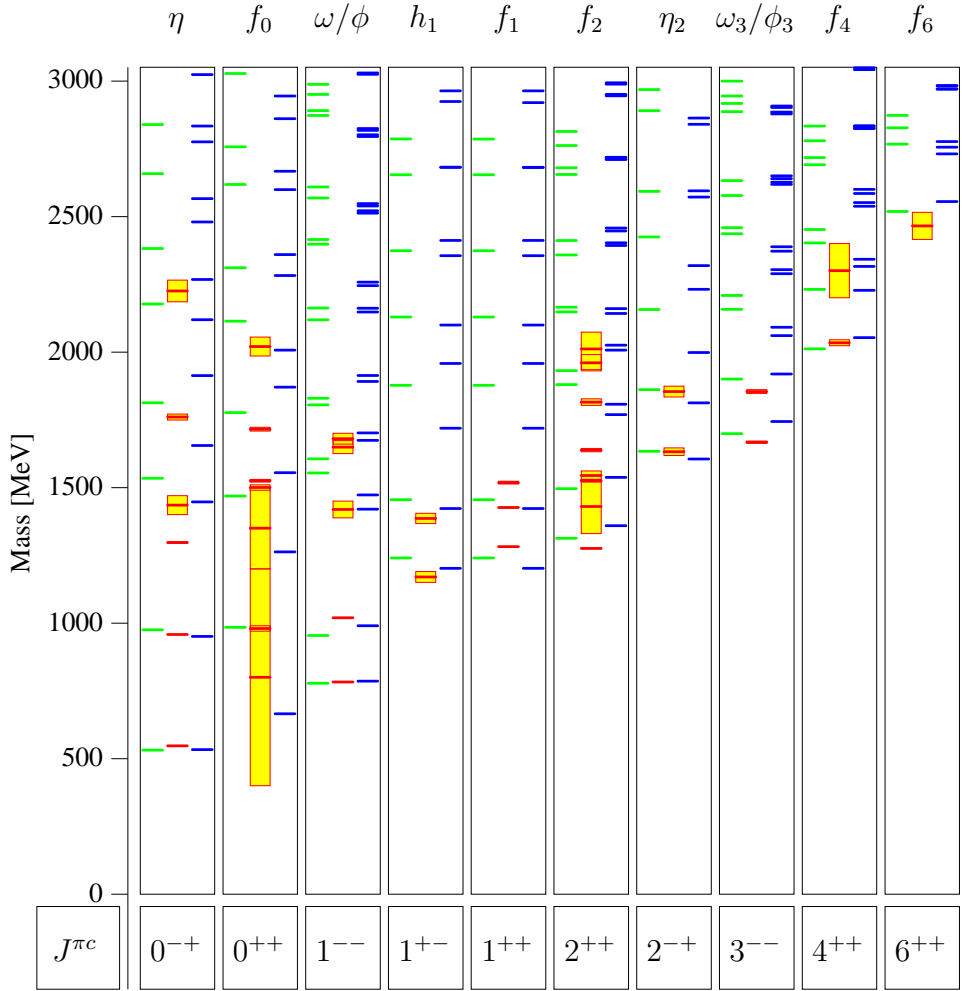}
\newline
       \protect\caption{The spectrum of the light mesons with isospin $I=0$. Left
    column for each $J^{\pi c}$: model $\cal A$; middle column for
    each $J^{\pi c}$: experimental masses and their error bars marked
    by the shadowed rectangles from 
  the {\sc Particle Data Group} (see \protect\cite{PDG00}); right
  column for each $J^{\pi c}$: model $\cal B$. There are
  experimental hints that the $\eta(1295)$ (plotted above according to
  the PDG data) might not really exist, see \protect\cite{Suh}.}
       \protect\label{fig:IsoScalar}
   \end{center}
\end{figure}

The resulting parameter sets are shown in
table \ref{tab:Parameters}. Note that the numerical values for
the confinement offset and slope, the 't Hooft couplings and the
constituent quark masses are comparable for both models although they
differ significantly in their confinement spin structure. Let us now
briefly discuss the resulting spectra in the isovector, isodublet and
isoscalar sectors: 

\begin{itemize}

\item Both models yield excellent Regge trajectories $M^2\propto J$, see
fig. (\ref{fig:Reggeplot}) for the light isovector mesons up to
$J=6$. The complete spectrum with all its radial excitations for
light mesons with isospin $I=1$ is shown in
fig. (\ref{fig:IsoVector}) and table \ref{tab:MassesIsoVector}. 

For the $a_0(980)$, we observe a remarkable
difference between the models $\cal A$ and $\cal B$: in contrast to
model $\cal A$, the spin
structure of the parameter set $\cal B$ obviously allows an interpretation
of the $a_0$ meson as a $q\bar q$ state. This is somewhat puzzling
since this state is often considered to be a $K\bar K$ molecule --- an
assignment that would be consistent with the fact that we indeed can not
describe the $a_0(980)$ with the parameters of model $\cal A$. We
refer to \cite{KlemptMuenz} for a detailed discussion
of the scalar meson spectrum and its interpretation in the framework
of the parameter set $\cal A$.

\item The complete kaonic spectrum is shown in
fig. (\ref{fig:IsoDublett}) and table \ref{tab:MassesIsoDublett}. 
We observe a very good agreement
with the established experimental data for angular momenta from $J=0$ to
$J=5$ with one exception: as in the isovector sector, the
$0^+$ state is lowered significantly in model $\cal B$ compared to
model $\cal A$. Let us briefly comment on this $K^*_0$ ground state:
compared to the PDG value of $M_{K^*_0}\approx 1430$MeV (see \cite{PDG00}), we find a
significant lowering of the mass in model $\cal B$ compared to model
$\cal A$. Indeed, there is an indication from a recent $K$--matrix
analysis for the scalar nonet 
that the $IP^\pi=\frac{1}{2}0^+$ ground state actually might
be around 1200MeV (see \cite{Anisovich}). Although there are good
reasons to compare our results to the ``bare poles'' of such an
analysis where the effects of decay--channel couplings are at least
partially taken into account, we have to refrain from a detailed
discussion until such an analysis has indeed been performed in all
meson sectors.

Furthermore, we note that the calculated states in the $K_1$ 
spectrum are each twice degenerate; the present interactions do not
distinguish the $S=0$ and the $S=1$ states and 
therefore do not show the experimentally observed splitting. However,
we want to mention that an alternative procedure for the
regularization of 't Hooft's instanton--induced force seperates these
two states; the reason is that the strict $J=0$ 
selection rule for this interaction is relaxed if the regularization
is applied before evaluating the occuring matrix elements. This effect,
although strongly suppressed, yields the correct splittings in the $K_1$
sector, but we will not discuss further the implications of this
slightly modified approach for the residual interaction in our model.

In the PDG listings (see \protect\cite{PDG00}), it is stated that the masses of the
$K_3$ and the $K_4$ need confirmation; they are therefore omitted in
the summary tables. We stress that our calculated masses for these
states fit perfectly in a linear Regge plot for the $K_J$ mesons 
--- in contrast to the experimental $K_3$ and $K_4$ data shown in
fig. (\ref{fig:IsoDublett}). We thus support the statement that these
have to be considered with caution. 

\item In fig. (\ref{fig:IsoScalar}) and table \ref{tab:MassesIsoScalar}, 
the results for the isoscalar mass
spectra are presented and compared to experimental data. Similar to
the pattern in the isovector and isodublet sector, we find
a remarkable downward shift of the scalar states in model $\cal B$
compared to model $\cal A$. We thus arrive at two alternative
interpretations of the scalar isoscalar $q\bar q$ spectrum: in model
$\cal A$, we might identify the lowest calculated $IJ^{\pi c}=00^{++}$ state
(mainly flavour singlet) either with the broad structure
$f_0(400-1200)$ or with the $f_0(980)$ meson and the second state at
$\approx$ 1500MeV as a member of the flavour octet; accordingly, this would leave
either the $f_0(980)$ or the $f_0(400-1200)$ and the $f_0(1370)$ as non--$q\bar q$
state. Let us note that if we furthermore regard the $\gamma\gamma$
decay width and the strong decay width of our lowest $f_0$ state in
model $\cal A$, we would slightly prefer the $f_0(980)$ to be the
$q\bar q$ candidate (see \cite{KlemptMuenz} and \cite{Ricken2}). 
In model $\cal B$, we do account for a very low--lying scalar
state (``$\sigma$ meson''), the next excitations to be identified with
states around 1250MeV and 1550MeV. As mentioned above, model $\cal B$
also accounts for the $a_0(980)$ meson as a $q\bar q$ state whereas this
can not be confirmed with the parameters of model $\cal A$. Clearly,
on the basis of the spectrum alone this interpretation can be
prelimininary at best, especially in view of the complexity in this
sector that arises from strong decay channel couplings and possible
mixtures with gluonic or other exotic states. A more detailed
discussion which also includes numerical results on the strong decay
widths of these states will be given in a forthcoming paper (see
\cite{Ricken2}). 

Finally we want to comment on the $\eta(1295)$: neither in
model $\cal A$ nor in model $\cal B$ we can interpret this as a
$q\bar q$ bound state. Indeed, it has recently been stated in \cite{Suh}
that there is no experimental evidence for an $\eta(1295)$ in the
reaction $p\bar p\to\pi^+\pi^-\pi^+\pi^-\eta$.

\end{itemize}

In summary, we find a good overall agreement
both in model $\cal A$ and $\cal B$. The
discrepancies, occuring {\it e.g.} in the $K_3$ or $K_4$ masses, can mostly
be traced back to questionable $q\bar q$ assignments of the considered
states. We find the right behaviour with respect to the linear Regge
trajectories $M^2\propto J$ up to highest angular momenta as has been shown in
fig. (\ref{fig:Reggeplot}).  Furthermore, due to the
instanton--induced effects our model shows the correct
splitting in the pseudoscalar sector independent of the underlying
parameter set. The effects of the different Dirac structures of the
confinement force in model $\cal A$ and $\cal B$ and their
implications for the non--relativistic reduction of the Salpeter
equation will be discussed in the detailed analysis of ref. \cite{Ricken1}.

Altogether, a linearly rising confinement potential
plus a residual interaction \`a la 't Hooft provides 
a very satisfacting description of the light $q \bar q$
spectra. Therefore we consider our approach based on the
Bethe--Salpeter equation in its instantaneous approximation as a
trustworthy framework for studying not only meson masses but also
their characteristic decays, especially in the pseudoscalar sector.

\section{Meson Decay Properties}      
                                        \label{DecayProperties} 

In previous publications (see \cite{MuenzPetry2}, \cite{MuenzPetry3},
\cite{GierscheMuenz} 
and \cite{Muenz}), we have studied various mesonic decay properties
such as the widths of the decays $\pi^0,\eta, \eta '\to\gamma\gamma$
or the decay constants $f_\pi$ and $f_K$. Furthermore, electromagnetic
form factors of $\pi$ and $K$ mesons have been investigated as well as
electromagnetic decays like $\rho\to\pi\gamma$ and related processes. We note that
instanton--induced vertices in strong decays were also studied (see
\cite{Ritter}); a more extensive publication in this context including quark loop
contributions is in preparation \cite{Ricken2}.  

We want to extend these approaches to other processes where $q\bar q$
bound states are involved. Firstly, we will pick up the recent discussion of
the pseudoscalar decay constants (see  \cite{FeldmannKrollStech1} and
\cite{FeldmannKrollStech2}) and quote our results for the $\eta$
and the $\eta '$ meson for their {\sl nonstrange} and {\sl strange} content 
separately. We will review the two photon decay of $J^\pi=0^-$
mesons and the related transition form factors at various photon
virtualities in section \ref{GammaGammaDecay} where our results for
$-q_i^2\to\infty$ will be linked to the pseudoscalar decay constants. 
We present the electromagnetic form factors of the charged $\pi$ and
$K$ mesons for the sake of completeness since they have not been
published yet in the parameter sets that we adopt in this paper.
For the same reason, we also present new results on the electromagnetic decay widths of the
processes ${\cal M}\to{\cal M}'\gamma$ that have already been studied
in ref. \cite{MuenzPetry3}.
Then, we will comment on the weak decays $\pi^+/K^+\to e^+\nu _e\gamma$
before we finally investigate our results on the form factors of 
the so--called $K_{\ell 3}$ decay.

Let us stress that these calculations concerning meson decay properties
are done in the parameter sets $\cal A$ and $\cal B$ that were
completely fixed with regard to the experimental meson mass spectra. We do
therefore not introduce new parameters or alter the models discussed
in the previous section since we aim at a global description of meson
masses as well as their characteristic decays.

\subsection{Pseudoscalar Decay Constants}      
                                        \label{DecayConstants} 

In recent years, the question of pseudoscalar decay constants ---
especially those of the $\eta$ and $\eta '$ mesons --- was discussed in
numerous publications (see {\it e.g.} \cite{FeldmannKrollStech1},
\cite{FeldmannKrollStech2}, and references therein). 
It has become clear that due to the mixing in the isoscalar sector,
the contributions of the {\sl nonstrange} and the {\sl strange} content (or between
singlet and octet contributions in an alternate mixing scheme) have to
be distinguished.

Let us start our discussion with the definition of the pseudoscalar
decay constant with the axial current $A^\pi_\mu=\bar
u\gamma_\mu\gamma_5 d$ for a pion with positive charge:
\begin{equation}
 \label{DecayConstant}
  \langle 0 | A ^\pi_\mu | \pi^+(p)\rangle = i p_\mu f_\pi   \quad .
\end{equation}
In a similar way, one can define the decay constant of the
$K^\pm$ meson; these constants are related with the
corresponding decay constants for the neutral mesons via $f_{{\cal
M}^0}=f_{{\cal M}^{\pm}} / {\sqrt 2}$ for ${\cal M}=\pi,K$. 

Due to the instantaneous approximation, $f_\pi$ can
be derived from the pion's Salpeter amplitude $\Phi _{(\pi)}(\vec p)$ in
the rest frame of the meson: 
\begin{eqnarray}
\label{f-pi-Salpeter}
  f_\pi &=& \frac{\sqrt 3}{M_\pi}\int\frac{d^4 p}{(2\pi)^4} \mbox{tr}
  \left[S_1^F(\frac P 2 + p)\Gamma ^P_{(\pi)}( p)S_2^F(-\frac P 2 +
  p)\gamma_0\gamma_5\right]\nonumber\\
&\stackrel{\scriptscriptstyle P=(M,\vec 0)}{=}&  \frac{\sqrt
  3}{M_\pi}\int\frac{d^3 p}{(2\pi)^3} \mbox{tr} 
  \left[\Phi _{(\pi)}(\vec p)\gamma_0\gamma_5\right]
\end{eqnarray}
where the trace is evaluated only on Dirac indices. It can be shown
analytically that the latter expression is proportional to the
difference of upper and lower component of the equal--time wave
function of the pion (see \cite{MuenzPetry1}). Therefore, $f_\pi$ is highly sensitive on
relativistic ingredients of the model in which it is calculated; in
non--relativistic models, the evaluation fails typically by orders of
magnitude. As has been shown in refs. \cite{MuenzPetry1} and \cite{MuenzPetry2}, the
electromagnetic decay widths for $\rho,\omega,\phi\to e^+e^-$ can be
treated in a similar way; the numerical results for the parameter sets
$\cal A$ and $\cal B$ of this work are comparable to the widths of
model V2 in these previous publications.

In table \ref{tab:DecayConstants}, the results for $f_\pi$ and $f_K$
are shown for the parameters of model $\cal A$ and $\cal
B$; obviously, we overestimate these observables by a factor of
$\approx$ 1.5 which is typical for
models with large constituent quark masses as found here, see
\cite{MuenzPetry2}. Although our
relativistic ansatz improves dramatically the 
result compared to non--relativistic calculations, we find in these
results a shortcoming of our model which might be related to the
instantaneous approximation of the underlying Bethe--Salpeter equation
and/or our refraining from ``running quark masses''.

Now let us focus the $\eta$ and the $\eta '$ meson, respectively. As
we have already emphasized, mixing between the {\sl nonstrange} and the {\sl strange} sector is
crucial for the understanding of the isoscalar states. We 
adopt the flavour decomposition
\begin{eqnarray}
\label{etaMixing}
\begin{array}{ccccc}
  |\eta\rangle &=&  X _\eta|n\rangle &+& Y _\eta|s\rangle \\
  |\eta '\rangle &=& X _{\eta '}|n\rangle &+& Y _{\eta '}|s\rangle 
\end{array}
\end{eqnarray}
where $|n\rangle := (|u\bar u\rangle + |d\bar d\rangle)/\sqrt 2$ and
$|s\rangle := |s\bar s\rangle$; the
coefficients obey $X_{\cal M}^2+Y_{\cal M}^2=1$ (${\cal
M}=\eta,\etap$). In table \ref{tab:XYCoefficients}, we compare the
coefficients, extracted from our calculated Salpeter amplitudes, with
experimental numbers from the {\sc Mark III} (see \cite{Mark3}) and
the {\sc DM2} (see \cite{DM2}) collaborations. The coefficients $X_\eta$
and $Y_\eta$ in model $\cal A$ and model ${\cal B}$ are consistent with
the findings of the experimental analysis whereas the $\eta'$ mixing
seems to be described not that well in model $\cal A$. Note that there are
also some new estimates for these 
coefficients based on a phenomenological analysis of the availiable
world data. They are also quoted in
table \ref{tab:XYCoefficients} and show that the experimental
determination might be not unique in a strict sense. However, it
should be added that these new results are found in the so--called
one--angle mixing scheme with
\begin{eqnarray}
\label{OneAngleMixing}
\left(
\begin{array}{c} 
\left| \eta \right \rangle \\ \left| \eta '\right \rangle
\end{array}
\right) = 
\left(
\begin{array}{cc} 
 \cos\phi & -\sin\phi \\ \sin\phi & \cos\phi
\end{array}
\right) 
\left(
\begin{array}{c} 
\left| n\right \rangle \\ \left|s \right \rangle
\end{array}
\right)
\end{eqnarray}
where by definition $|X_{\eta }|=|Y_{\eta '}|$ and $|Y_{\eta }|=|X_{\eta
'}|$ holds so that the numerical values of
ref. \cite{FeldmannKrollStech1} and \cite{Cao} have to be considered
with care. Indeed we find that $\eta$ and $\eta '$ are not simply
related by a flavour rotation; in our calculation, the radial
amplitudes of these isoscalar states differ considerably.

Therefore, adopting the general mixing scheme of eq. (\ref{etaMixing}), we define
the decay constants for the $\eta$ and the $\eta '$ meson as follows:
\begin{equation}
 \label{etaDecayConstant}
  \langle 0 | A^j _\mu | {\cal M}(p)\rangle = i p_\mu f_{\cal M}^j
  \qquad(\;j=n,s\;; \quad{\cal M}=\eta,\etap\;)
\end{equation}
with $A^n_\mu=(\bar u\gamma _\mu\gamma _5 u + \bar d\gamma _\mu\gamma _5
d)/\sqrt 2$ and $A^s_\mu=\bar s\gamma _\mu\gamma _5 s$. In practise,
these constants are computed as in
eq. (\ref{f-pi-Salpeter}). In table \ref{tab:DecayConstants}, we present
our results using the parameters of the models $\cal A$ and $\cal
B$. We stress that the quoted results of \cite{FeldmannKrollStech1}
are found in the one--angle mixing scheme of eq.(\ref{OneAngleMixing}) so that a comparison with our
ansatz has to be done with some caution. We conclude that although $f_\pi$ and
$f_K$ are overestimated in our model, we find plausible results of
roughly the same quality for the $\eta$ and $\eta '$ decay constants compared to
\cite{FeldmannKrollStech1} due to the instanton--induced
mixing of these isoscalar states.

\subsection{Two Photon Decays}      
                                        \label{GammaGammaDecay} 

In this section, we will discuss the two photon decays of
pseudoscalar mesons not only for two real photons in the final state
($q_1^2 = q_2^2 = 0$) but also the $0^-$ transition form factors in the case
that either one ($q_1^2 = 0$, $q_2^2 \not = 0$) or both ($q_1^2 \not = 0$, $q_2^2
\not = 0$) of the photons are virtual.

In fig. (\ref{fig:TPD}), the $\gamma\gamma$ decay of pseudoscalar mesons is
shown, calculated in lowest order in the Mandelstam formalism (see \cite{Mandelstam}). The related
matrix element for the transition ${\cal M}\to\gamma\gamma$ with
${\cal M}=\pi^0,\eta, \eta '$ reads
\begin{eqnarray}
\label{MatrixElement}
  T^{\cal M}_{\gamma\gamma}(q_1, q_2) = i\sqrt 3\int\frac{d^4 p}{(2\pi)^4}
  \mbox{tr}&\bigg[& S^F(\frac P 2 + p)\Gamma ^P_{({\cal M})}(p)S^F(-\frac P 2 + p)
  \not\!\varepsilon _2 \,S^F(\frac P 2 + p - q_1)\not\!\varepsilon _1 \\ &+
  &S^F(\frac P 2 + p)\Gamma^P_{({\cal M})}(p)S^F(-\frac P 2 + p)\not\!\varepsilon _1
  \, S^F(\frac P 2 + p - q_2)\not\!\varepsilon _2 \bigg]
   \quad ,\nonumber
\end{eqnarray}
where $\Gamma^P_{({\cal M})} (p)$ is the vertex function for the
pseudoscalar meson $\cal M$ defined in
eq. (\ref{VertexFunction}). The polarization vectors
$\epsi _i$ obey $\epsi _i^2=-1$ and $\epsi _i\cdot k_i=0$ ($i=1,2$)
for real photons. Here, the factor $\sqrt 3$ originates from the trace
over colour indices; the trace in eq. (\ref{MatrixElement}) is
understood with respect to 
Dirac and flavour indices.

From Lorentz invariance, it is possible to derive another expression for this
matrix element for pseudoscalar mesons
\begin{equation}
\label{TandF}
T^{\cal M}_{\gamma\gamma}(q_1, q_2)= \alpha\: \epsilon 
  _{\mu\nu\alpha\beta} \:\varepsilon ^\mu _1(q_1)\varepsilon ^\nu
  _2(q_2)q_1^\alpha q_2^\beta \cdot F^{\cal M}_{\gamma\gamma}(q^2_1, q^2_2) 
\end{equation}
in terms of the transition form factor $F^{\cal M}_{\gamma\gamma}(q^2_1,
q^2_2)$; here, $\alpha$ is the fine structure constant.
With this last equation, it is now possible to relate our matrix element
calculated via eq. (\ref{MatrixElement}) with the transition form factor and
the decay width defined by
\begin{eqnarray}
\label{GammaGammaWidth}
\Gamma^{\cal M}_{\gamma\gamma}(Q^2_1, Q^2_2) = \alpha
 ^2\frac{M_{\cal M}^3}{64\pi}\left|F^{\cal M}_{\gamma\gamma}(Q^2_1, Q^2_2)\right|^2 
 \qquad \mbox{with}\quad Q_i^2 := - q_i^2 \quad .
\end{eqnarray}

It is worth noting  that in a similar way it is possible to calculate the two
photon widths $\Gamma _{\gamma\gamma}$ in our relativistic quark model 
not only for $J^\pi=0^-$ mesons but
also for $J^\pi=0^+, 2^\pm, 4^\pm\ldots$ mesons (including $c\bar c$ and $b\bar
b$ bound states with an additional one--gluon--exchange potential) and
their radial excitations in  reasonable agreement with the (rare)
experimental data (see \cite{Muenz}). 

\begin{figure}
  \protect\begin{center}
    \leavevmode
       \protect\input{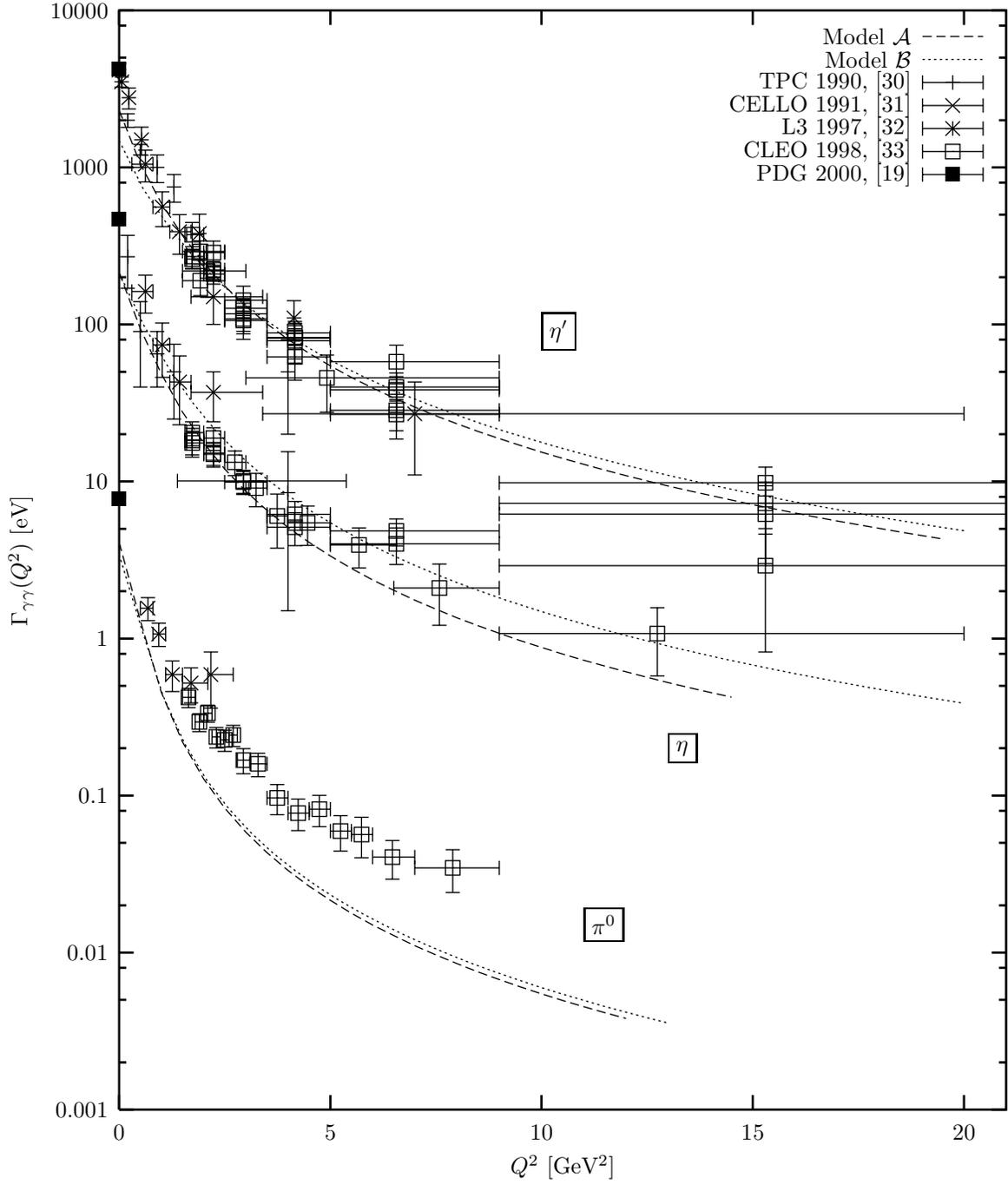}
\newline
       \protect\caption{The decay widths for the processes
  $\pi^0,\eta,\etap\to\gamma\gamma^*$ as a function of the momentum
  transfer of the virtual photon, calculated with
  the parameters of model $\cal A$ and model $\cal B$.}
      \protect \label{fig:TwoPhotonDecayWidths}
   \end{center}
\end{figure}

In table \ref{tab:GammaGammaWidths}, we show our results for the
widths of the decays $\pi^0,\eta,\etap\to\gamma\gamma$ into two real
photons; obviously, we underestimate these widths and as for the
pseudoscalar decay constants, we find a discrepancy of about a factor of
1.5 in the amplitude. Indeed, this can be
understood with the results of the foregoing subsection since --- {\it e.g.} for
the pion --- the relation
$\Gamma^{ \pi^0}_{\gamma\gamma}(0,0)\propto 1/f_\pi^2$ holds. From
fig. (\ref{fig:TwoPhotonDecayWidths}), one observes that for all $Q^2$
the two photon width $\Gamma^{\pi^0}_{\gamma\gamma}(Q^2):=
\Gamma^{\pi^0}_{\gamma\gamma}(Q^2:=Q^2_1, 0)$ is underestimated. We conclude
that for a deeply bound particle such as the pion the instantaneous
approximation shows up its shortcomings although the meson Salpeter
amplitudes are correctly boosted. We also present the widths for the
scalar meson decays $f_0(400-1200),f_0(980),a_0(980)\to\gamma\gamma$
in table \ref{tab:GammaGammaWidths}. Note that each of these decays
can only be calculated in one of the parameter sets $\cal A$ or $\cal
B$ due to the different $q\bar q$ assignments to the mesons in the
scalar sector in both models, see sect. \ref{ParametersSpectra}.

Let us now study these results in some more detail. In
ref. \cite{Brodsky81}, Brodsky and Lepage presented the
well--known and parameter--free interpolation formula for the pion
transition form factor:  
\begin{equation}
  \label{OnePoleInterpolation}
  F^{\pi^0}_{\gamma\gamma}(Q^2,0) = \frac{6 C_\pi f_\pi}{Q^2 + 4\pi^2f_\pi^2}
\end{equation}
with the charge factor $C_\pi:=1/(3\sqrt 2)$ coming from the quark
flavours. Although Brodsky {\it et al.} recently proposed a slightly modified version of
this formula (see \cite{Brodsky98}), it works quite well in its original version 
and leads to the famous limit 
\begin{equation}
  \label{OnePoleLimit}
\lim_{Q^2\to \infty}  Q^2 F^{\pi^0}_{\gamma\gamma}(Q^2,0) =  6 C_\pi
f_{\pi} = 2 f_{\pi^0}
\end{equation}
for the form factor of the decay $\pi^0\to\gamma\gamma ^*$ in the
asymptotic region; note that here $f_{\pi^0}:=f_\pi/\sqrt{2}\approx
93$MeV. On the other hand, we recover
$\Gamma^{\pi^0}_{\gamma\gamma}(Q^2\to 0,0)\propto 1/f_\pi^2$ with
eqs. (\ref{GammaGammaWidth}) and (\ref{OnePoleInterpolation}) for real
photons as quoted above.

For the $\eta$ and the $\etap$ , the situation is somewhat
different. Since these isoscalar mesons are mixed {\sl
nonstrange/strange} states, one has to
distinguish between the contributions of different flavour structures,
see eq. (\ref{etaMixing}). In \cite{FeldmannKroll}, the authors
propose a two--pole interpolation formula for the $\eta/\etap$
transition form factors yielding
\begin{equation}
\label{TwoPoleLimit}
\lim_{Q^2\to \infty}  Q^2 F^{\cal M}_{\gamma\gamma}(Q^2,0) = 6 C_n f_{\cal M}^n +
           6 C_s f_{\cal M}^s    
\end{equation}
for ${\cal M}=\eta,\etap$ in the asymptotic limit. Here, $f_{\cal M}^j$ with $j=n,s$ are the
decay constants defined in eq.(\ref{etaDecayConstant}) and
$C_n=5/(9\sqrt 2)$ and $C_s=1/9$ are charge factors coming from the
{\sl nonstrange}/{\sl strange} quark flavours. If we compare our
numerical results for the
left--hand side of eq. (\ref{TwoPoleLimit}) and for the right--hand
side from the values of
table \ref{tab:DecayConstants}, we find that we do not obtain the right
limit as $ Q^2\to\infty$ for the transition form
factors ({\it e.g.} $Q^2 F^{\pi^0}_{\gamma\gamma}(Q^2\to\infty,0)\approx
70$MeV but $6 C_\pi f_{\pi}\approx 150$MeV in both models). We want to
stress that nevertheless we find a good agreement with the 
experimental results for $Q^2 F^{\cal M}_{\gamma\gamma}(Q^2)$ or $\Gamma^{\cal
M}_{\gamma\gamma} (Q^2)$, respectively, for ${\cal M}=\eta,\eta '$ --- see
fig. (\ref{fig:TwoPhotonDecayWidths}). 

Finally, we want to comment on the decay of a $J^\pi=0^-$ meson into
two virtual photons. We will focus on the case of identical virtuality
of the outgoing photons ($Q_1^2=Q_2^2=:Q^2$) but we want 
to emphasize that in general the whole $Q_1^2$--$Q_2^2$ plane can be
calculated in our model. A result for $Q^2\to\infty$ from operator
product expansion yields 
\begin{equation}
  \label{OnePoleLimitVV}
\lim_{Q^2\to \infty}   Q^2 F^{\pi^0}_{\gamma\gamma}(Q^2,Q^2) = 2  C_\pi f_{\pi}
\end{equation}
for the transition form factor of the pion (see \cite{Novikov} and
\cite{Anselm}). We can confirm this result in our  
calculations in contrast to the limits of the foregoing discussion since now
the virtuality distribution of the two outgoing photons is symmetric even at
very large $Q^2$. Furthermore, we can give a similar relation for the
form factor of the decays $\eta,\etap\to\gamma^*\gamma^*$  in the
limit of large $Q^2$ from analytical considerations in the framework
of our model (see the Appendix for details):
\begin{equation}
  \label{TwoPoleLimitVV}
\lim_{Q^2\to \infty}   Q^2 F^{\cal M}_{\gamma\gamma}(Q^2,Q^2)= 2 C_n f_{\cal M}^n + 2 C_s f_{\cal M}^s 
\end{equation}
for ${\cal M}=\eta,\etap$; the charge factors $C_j$ and the decay constants $f_{\cal M}^j$
($j=n,s$) are defined as above. In
figs. (\ref{fig:VirtualVirtualModelA}) and
(\ref{fig:VirtualVirtualModelB}), these form
factors are plotted together with their limits for large $Q^2$ with
model $\cal A$ and $\cal B$; no experimental data exist in this
kinematical region so far. Obviously, the numerically evaluated form
factors in both models indeed show up the limits that we found in our analytical
calculations, see the Appendix.

\subsection{The Electromagnetic Form Factors of the Charged $\pi$ and
$K$ Mesons}      
                                        \label{piKFormFactors} 

For the sake of completeness, we want to present the electromagnetic
form factors of the $\pi^\pm$ and the $K^\pm$ in this section since
they have not been published in the models $\cal A$ and $\cal B$ up to
now. 

The meson form factors for the transitions
$\pi^\pm(P)\to\pi^\pm(P')\gamma^*(q)$ and $K^\pm(P)\to K^\pm(P')\gamma^*(q)$
with a (spacelike) photon virtuality $q^2=(P-P')^2=:-Q^2<0$ are
defined as follows:
\begin{equation}
\label{FormFactorDef}
\left\langle {\cal M} (P')\left|J_\mu(0)\right| {\cal M}(P)\right\rangle = {\cal Q} F_{\cal
M}(Q^2)
\left(P+P'\right)_\mu 
\end{equation}
Here, $P(P')$ is the four--momentum of the incoming (outgoing) meson
${\cal M}= \pi^\pm,K^\pm$; the factor ${\cal Q}=e_1+e_2$ denotes the
meson charge. As has been shown in 
ref. \cite{MuenzPetry3}, we derive this matrix element in the
Bethe--Salpeter approach via the Mandelstam formalism (see
\cite{Mandelstam}) on the basis of a 5--point Greens function yielding
\begin{eqnarray}
\label{FormFactorBSE}
{
\left\langle{\cal M} (P') \left |\:j^\mu (0)  \right |{\cal M}( P)
\right\rangle}
= - \:\int\frac{d^4 p}{(2\pi)^4} \:\mbox{tr}\bigg[ &e_1 &\bar{\Gamma}_{({\cal M})}
^{P^\prime}(p-\frac q 2)S_1^F( \frac P 2 +p - q) \gamma^\mu S_1^F(
\frac P 2 +p)\Gamma ^P_{({\cal M})}(p)S_2^F(-\frac P 2 +p)    \\ 
+& e_2 &\bar{\Gamma}_{({\cal M})}
^{P^\prime}(p+\frac q 2)S_1^F( \frac P 2 +p)\Gamma_{({\cal M})} 
^P(p)S_2^F(-\frac P 2 +p) \gamma^\mu S_2^F(-\frac P 2 +p+q)\bigg]
\qquad , \nonumber
\end{eqnarray}
where the trace is understood with respect to the Dirac indices. The details of this
procedure as well as a discussion of the shape of the form factor
$F_\pi(Q^2)$ at very low $Q^2$ can be found in \cite{MuenzPetry3}.

\begin{figure}
  \protect\begin{center}
    \leavevmode
       \protect\input{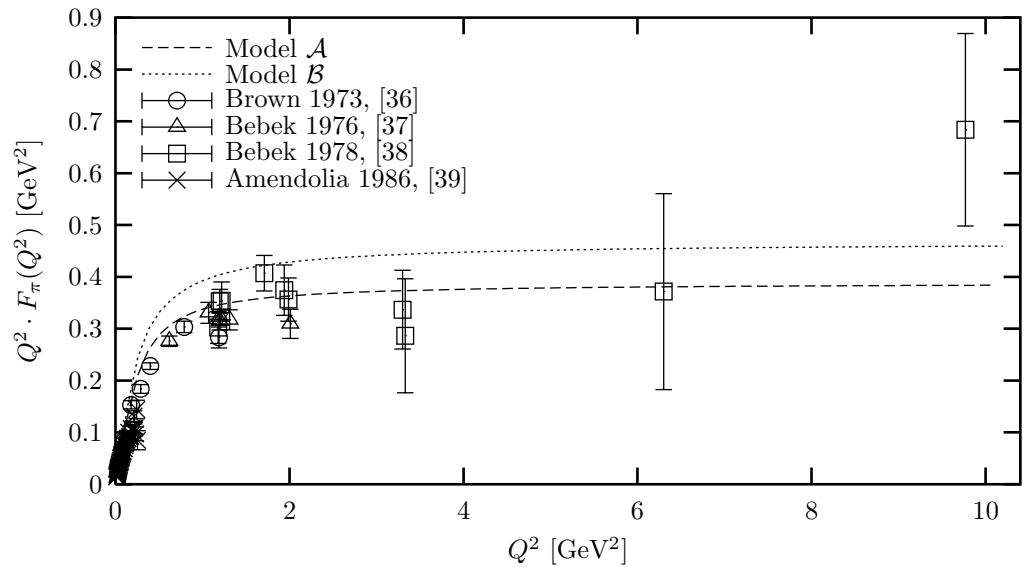}
\newline
       \protect\caption{The electromagnetic form factor $Q^2\cdot F_\pi(Q^2)$
  of the charged pion, calculated with the parameters of model
  $\cal A$ and model $\cal B$. Note that the correct shape of the form
  factor beyond $\approx 1$GeV${}^2$ can be traced back to the
  application of the full Lorentz boost, see \protect\cite{MuenzPetry3}.} 
      \protect \label{fig:pi_FormFactor}
   \end{center}
\end{figure}

\begin{figure}
  \protect\begin{center}
    \leavevmode
       \protect\input{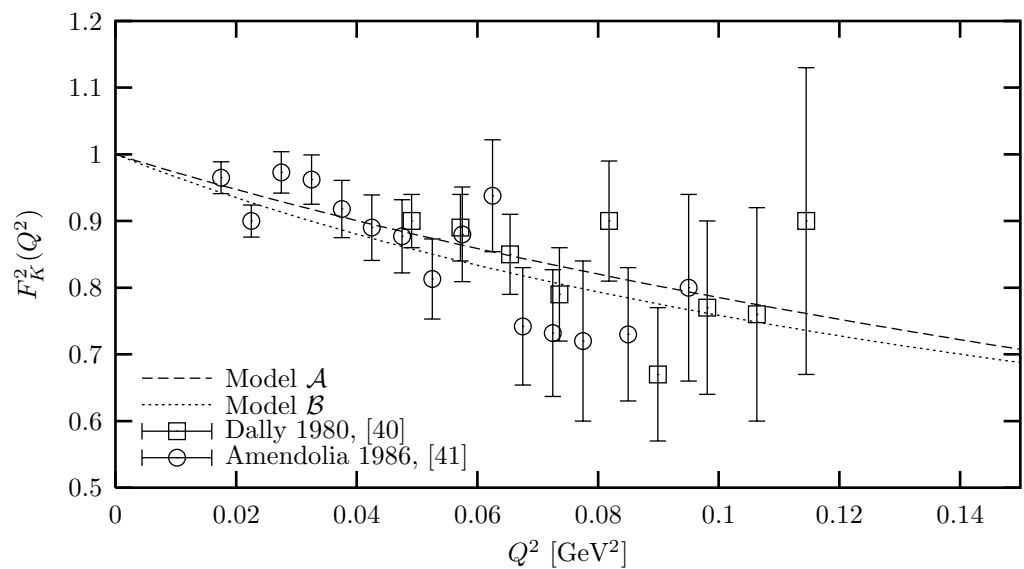}
\newline
       \protect\caption{The electromagnetic form factor $F^2_K(Q^2)$
  of the charged kaon, calculated with the parameters of model
  $\cal A$ and model $\cal B$.} 
      \protect \label{fig:K_FormFactor}
   \end{center}
\end{figure}

We show the form factor $Q^2\cdot F_\pi(Q^2)$ in
fig. (\ref{fig:pi_FormFactor}). The correct behaviour of the form factor at
very large $Q^2$ can be traced back to the Lorentz boost that is
applied to the outgoing vertex function. The charged kaon form factor
is plotted in fig. (\ref{fig:K_FormFactor}); we conclude that our model
provides a satisfying description of the electromagnetic
$\pi^\pm$ and $K^\pm$ form factors in both parameter sets. Let us
finally note that the correct form factor normalization $F_{\cal
M}(Q^2=0)=1$ is a consequence of the normalization condition of the
Bethe--Salpeter equation; we therefore do not need to impose an {\it
ad hoc} normalization of the electromagnetic form factor.

\subsection{The Electromagnetic Decay Widths for ${\cal M}\to{\cal M}'\gamma$}      
                                        \label{EMDecayWidths} 

To complete our discussion on electromagnetic processes involving
mesons and to update the results of ref. \cite{MuenzPetry3}, we will
briefly comment on the widths of the decays ${\cal M}(P)\to{\cal M}'(P')\gamma(q)$.
For this purpose, we extend the matrix element in eq. (\ref{FormFactorBSE}) to
processes like $\rho\to\pi\gamma$ where also $J\not=0$ mesons
occur. The related decay form factor $F_{\rho\pi}(Q^2)$ with
$Q^2=-q^2$ is then given by
\begin{equation}
\label{EMDecayDef}
\left\langle \pi (P')\left|J_\mu(0)\right| \rho(P,\lambda)\right\rangle =
{\cal Q} F_{\rho\pi}(Q^2)
\varepsilon_{\mu\nu\alpha\beta}\frac{\epsilon^\nu_\rho(
P,\lambda)P^{\prime\alpha}P^\beta}{M_\rho}  \quad .
\end{equation}
Here, $\epsilon _\rho(P,\lambda)$ denotes the polarization vector of the
$\rho$ meson with spin projection $\lambda$; analogue definitions hold for
all processes that will be discussed in the following. The general
decay width can be computed via  
\begin{equation}
\label{EMWidths}
\Gamma _{{\cal M}\to{\cal M}'\gamma} = \alpha
\frac{1}{2J+1}\frac{q}{M^2_{\cal M}} \sum_{M_J M_{J'}}\left| \epsilon
^\mu_\gamma(\vec q,+1)\left\langle
{\cal M}' (P',J',M_{J'})\left|J_\mu(0)\right| {\cal M}(P,J,M_J)\right\rangle \right|^2
\end{equation}
where $\epsilon_\gamma$ is the polarization vector of the photon with
three--momentum $\vec q=q\vec e_z$ and the matrix element is evaluated
in the rest frame of meson $\cal M$ with $P=(M,\vec 0)$.

We show the widths for various meson decays in
table \ref{tab:EMDecayWidths} and compare them to the latest PDG data
compilation (see \cite{PDG00}). As already stated in
\cite{MuenzPetry3}, we clearly underestimate the processes with a pion in the
final state, especially in model $\cal B$. Obviously this is again a significant shortcoming of the
instantaneous approximation for deeply bound particles such as the
pion; the resulting lack of retardation effects seems to spoil the correct overlapping of
the associated wave functions in our calculation. Note however for the
$\rho/\omega\to\pi\gamma$ decays that an 
exact SU(2) flavour symmetry --- {\it i.e.} $m_u=m_d$ --- implies
$\Gamma_{\rho^\pm\to\pi^\pm\gamma}=\Gamma_{\rho^0\to\pi^0\gamma}=
\frac{1}{9}\Gamma_{\omega^0\to\pi^0\gamma}$ as can be seen from the
calculated values; therefore the results for the experimental widths
in table \ref{tab:EMDecayWidths} are rather puzzling.  

We stress that the excellent results for $\Gamma_{{\cal M}\to\eta\gamma}$
with ${\cal M}=\rho^0,\omega,\phi$ indicate that the coefficients
$X_\eta$ and $Y_\eta$ for the {\sl nonstrange/strange} flavour mixing
are determined well in both models, see
table \ref{tab:XYCoefficients}. In addition, we find plausible
results for the decays $\eta '\to\rho^0\gamma$ and $\eta
'\to\omega\gamma$ with the calculated widths in both models close to
the experimental findings.

Due to the $J=0$ selection rule for 't Hooft's flavour dependent
interaction, no flavour mixing is induced for the $\omega,\phi$ and $f_1$
mesons. It is therefore not surprising that we find vanishing decay
widths for some processes since {\it e.g.} a pure {\sl nonstrange}
state ($\sim n\bar n )$ like the $\omega$ 
does not couple to a pure {\sl strange} state ($\sim s\bar s$) like
the $\phi$ thus yielding $\Gamma_{\phi\to\omega\gamma}=0$ in both
models; we do not quote these zero widths in table \ref{tab:EMDecayWidths}. 

In the kaonic sector, we find a very good agreement of the
$\Gamma_{K^*\to K\gamma}$ widths in model $\cal A$ with the PDG values; our numerical
results in model $\cal B$ slightly underestimate the experimental data
although we can describe the correct ratio between the neutral and the
charged decay mode.

\begin{figure}
  \protect\begin{center}
    \leavevmode
       \protect\input{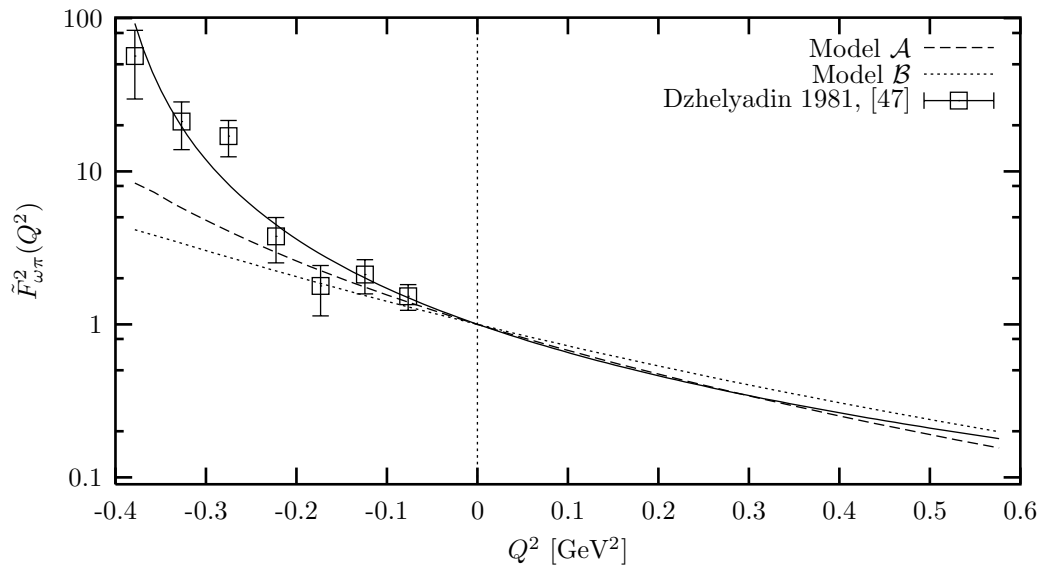}
\newline
       \protect\caption{The normalized decay form factor $\tilde F^2_{\omega\pi}(Q^2)=(F_{\omega\pi}(Q^2)/F_{\omega\pi}(0))^2$, calculated with the parameters of model
  $\cal A$ and model $\cal B$. The solid line is the pole fit of ref. \protect\cite{Dzhelyadin81} with the parameter $\Lambda=0.65$GeV, see eq. (\ref{FormFactorPoleFit}).} 
      \protect \label{fig:OmegaPi_FormFactor}
   \end{center}
\end{figure}

Let us now come back to the form factor defined in
eq. (\ref{EMDecayDef}). For the decay $\omega\to\pi^0\gamma^*$ with
the virtual photon decaying into $\mu^+\mu^-$, there are some
experimental values for the normalized form factor $\tilde
F_{\omega\pi}(Q^2)=F_{\omega\pi}(Q^2)/F_{\omega\pi}(0)$ at timelike
photon virtuality $q^2=-Q^2>0$. In
fig. (\ref{fig:OmegaPi_FormFactor}), we have plotted these experimental
data points and our numerical results, calculated with the parameters
of the models $\cal A$ and $\cal B$. Note that for $Q^2>-(m_q^2+m_{\bar
q}^2)$ our model will become ill--defined since we cannot guarantee
confinement for timelike momentum transfers; we therefore did not
compute $\tilde F_{\omega\pi}(Q^2)$ beyond the threshold at $\approx
-0.37$GeV${}^2$ of model $\cal A$. Comparing our calculated
curves with the experimental data and with a pole fit according to 
\begin{equation}
\label{FormFactorPoleFit}
\tilde F_{\omega\pi}(Q^2) = \frac{1}{1+Q^2/\Lambda ^2}
\end{equation}
we find that we underestimate the shape of the form factor in the
timelike region; obviously, this again can be traced back to the
shortcomings of our model in the case of a $\pi$ meson in the final
state as has been
discussed above. However, our results are comparable for $Q^2>0$ with
the pole fit extracted in the experimental study in
ref. \cite{Dzhelyadin81}. The authors found $\Lambda _{\mbox{\tiny
exp}}=(650\pm30)$MeV while our form factors would merely coincide with a
simple $\rho$ pole ansatz, {\it i.e.} $\Lambda _\rho\approx770$MeV; this
discrepancy becomes obvious for $Q^2<0$.

Summarizing this subsection on the electromagnetic decay widths, we
conclude that we find a good overall agreement with the experimental
data on the level of a factor of $\approx$ 1.5 in the amplitudes with
one exception: for a pion in the final state, our calculations fail
significantly.

\subsection{The Decays $\pi^+\to e^+\nu _e\gamma$ and $K^+\to e^+\nu _e\gamma$}      
                                        \label{piKDecay} 

The so--called $\pi_{\ell 2\gamma}$ decay, {\it i.e.} $\pi^+ (P) \to\ell^+ (p_\ell)
\nu _\ell(p_\nu)\gamma (q)$, with the lepton $\ell=e,\mu$ and the
analogously defined $K_{\ell 2\gamma}$ decay have been studied
extensively both experimentally and theoretically in the last
decade. For a muon in the final state, the 
$\pi_{\ell 2\gamma}$ decay would be dominated by Bremsstrahlung
effects; however, for $\ell=e$ this contribution is strongly helicity
suppressed such that the structure dependent parts of the related
amplitude can be measured. The matrix element for this process reads
\begin{equation}
\label{Mgamma}
{\cal M}_{\pi^+\to e^+\nu _e\gamma} = - \frac{eG_F}{\sqrt 2} V^*_{ud}
\varepsilon^{\mu *}(q) M^{\ell 2\gamma}_{\mu\nu}(p,q)\bar
u(p_\nu)\gamma^\nu(\Id+\gamma_5)v(p_e) 
\end{equation}
where $\bar u(p_\nu)$ and $v(p_e)$ denote the Dirac spinors for the
neutrino and the positron (see \cite{Bryman}, \cite{DonoghueHolstein}
and the PDG mini--review in \cite{PDG00}). Here, the outgoing photon
is real so that for its polarization vector $\varepsilon\cdot q=0$
with $q^2=0$ holds. The quantity $M^{\ell 2\gamma}_{\mu\nu}$ can be
formulated as the time--ordered product of the electromagnetic current
$J_\mu^{\mbox{\tiny el.magn.}}$ and the weak current
$J_\nu^{\mbox{\tiny weak}}$ and reads
explicitly:
\begin{eqnarray}
\label{Mmunu}
M^{\ell 2\gamma}_{\mu\nu}(p,q) &=& \int d^4 x\: \left\langle 0\left|
J_\mu^{\mbox{\tiny el.magn.}} (x) J_\nu^{\mbox{\tiny weak}} (0)
\right|\pi^+(P)\right\rangle  e^{iq\cdot x} \\
&=& f_\pi\left(g_{\mu\nu} -  \left\langle \pi^+ (P-q)
\left|J_\mu^{\mbox{\tiny el.magn.}} \right|\pi^+(P)\right\rangle
\frac {(P-q)_\nu}{(P-q)^2-M_\pi^2}\right)  \nonumber \\
&& -h_A\left(\left(P-q\right)_\mu q_\nu-
q\cdot\left(P-q\right)g_{\mu\nu}\right) +ih_V
\epsilon_{\mu\nu\alpha\beta}q^\alpha P^\beta  \nonumber
\end{eqnarray}
The first term including $f_\pi$ is the so--called Born part; the subscripts at
the form factors $h_V$ and $h_A$ denote their origin either in the
vector ($V$) or in the axial vector ($A$) part of the weak current.
Note that an additional axial form factor occurs if the outgoing
photon is virtual, {\it i.e.} for the decays $\pi^+/K^+\to e^+\nu _e e^+e^-$;
we will not study these processes here.

In eq. (\ref{Mmunu}), the Born terms contain the pion form factor;
this quantity has been calculated in the 
framework of our model, see the foregoing subsection. As it stands, the
full matrix element is gauge invariant due to the inclusion of the
Born terms. In the following, we will only consider the structure
dependent contributions, but we will take care that in the extraction of the
form factors from the full tensor $M^{\ell 2\gamma}_{\mu\nu}$ no terms
will occur that violate gauge invariance.

The structure dependent parts of eq. (\ref{Mmunu}) can be calculated
in the Mandelstam formalism. The result is analogous to the matrix element
of the two photon decay, see eq. (\ref{MatrixElement}) and
fig. (\ref{fig:TPD}) except that one photon line is substituted by a $W^+$
boson line with the typical $V-A$ Dirac structure and the flavour
matrix $\lambda _1 - i\lambda _2$. The calculation can be extended to
the $K_{\ell 2\gamma}$ decays by inserting kaon observables
$M_K,F_K$ in eq. (\ref{Mmunu}) and by using the corresponding matrix in
flavour space. 

In table \ref{tab:RadLept}, we present the results for the form
factors $h_V$ and $h_A$ of the $\pi_{\ell 2\gamma}$ decay as well
as for the $K_{\ell 2\gamma}$ decay. They are compared with the world
averages of the {\sc Particle Data Group} (see \cite{PDG00}). The form
factors of the $K_{\ell 2\gamma}$ decay are known only incompletely; we
quote the experimental results for the sum and the difference of $h_V$
and $h_A$. In the case of the $\pi_{e 2\gamma}$ decay, one has to
regard the values of the axial form factor $h_A$ with care. Since
usually only the ratio $\gamma:=h_A/h_V$ is measured and only two
direct determinations of $h_V$ are presented up to now, the PDG
results for $h_A$ are determined via this ratio with the input
$h_V^{\mbox{\tiny CVC}}=(0.0259\pm 0.0005)/M_\pi^2$ from the CVC
prediction. Despite this caveat, we will however quote the results of
ref. \cite{PDG00} for a comparison of our results with experimental
data. We obtain excellent results compared to the PDG averages in
both models, see table \ref{tab:RadLept}.

\subsection{Form Factors of the $K_{\ell 3}$ Decay}      
                                        \label{Kl3Decay} 

\begin{figure}[h]
  \protect\begin{center}
    \leavevmode
       \protect\input{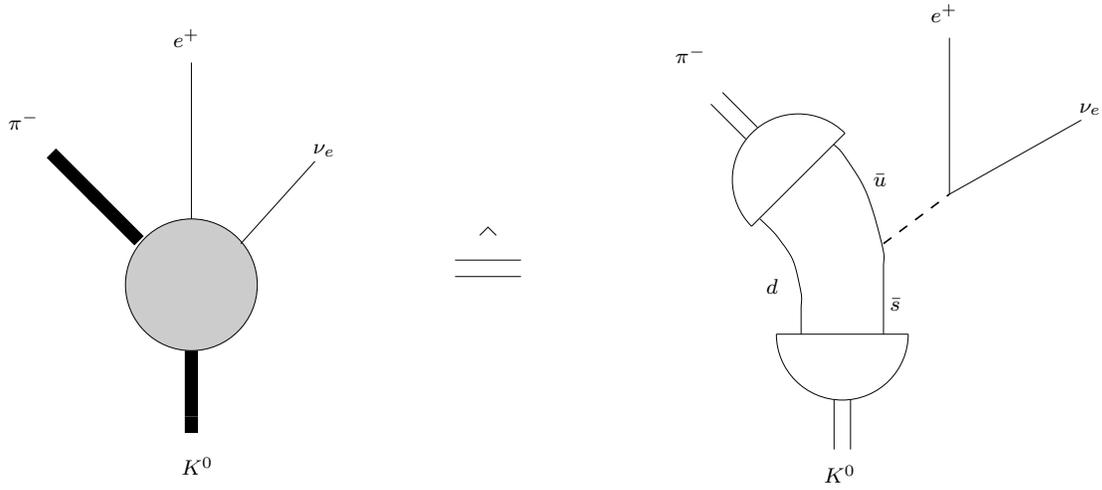}
\newline
       \protect\caption{The decay $K^0\to\pi^-\ell
        ^+\nu_\ell$ in lowest order in the Mandelstam formalism.}
       \protect\label{fig:Kl3Decay}
   \end{center}
\end{figure}

The processes $K^+\to\pi^0\ell ^+\nu_\ell$ and $K^0\to\pi^-\ell
^+\nu_\ell$ are called $K^+_{\ell 3}$ and $K^0_{\ell 3}$ decay,
respectively. They are usually parameterized in terms of two form factors for
which isospin invariance requires
$f_\pm^{K^+\pi^0}=f_\pm^{K^0\pi^-}/\sqrt{2}=:f_\pm$ (see the
minireview in \cite{PDG00}): 
\begin{eqnarray}
\left\langle\pi^0 (P')\left|J_\mu^{\mbox{\tiny
weak}}\right|K^+(P)\right\rangle &=& f_+^{K^+\pi^0}
\left(P+P'\right)_\mu +  f_-^{K^+\pi^0}
\left(P-P'\right)_\mu  \\
\left\langle\pi^- (P')\left|J_\mu^{\mbox{\tiny
weak}}\right|K^0(P)\right\rangle &=& f_+^{K^0\pi^-}
\left(P+P'\right)_\mu + f_-^{K^0\pi^-}
\left(P-P'\right)_\mu \qquad . \nonumber
\end{eqnarray}
In fig. (\ref{fig:Kl3Decay}), we show the $K^0_{\ell 3}$ decay in our
Bethe--Salpeter quark model picture. Note that we describe the
emission of the leptonic pair by a $W^+$ boson coupling to the {\sl strange}
quark and decaying into $\ell^+\nu_\ell$. 

For heavy quark systems, semileptonic decays of this type can also be
calculated in our model if either an additional one--gluon exchange
potential (see \cite{ZoellerHainzl}) or an appropriate extension of 't Hooft
instanton--induced force is adopted; a subsequent paper on this subject is currently
in preparation.

The matrix element for the $K_{\ell 3}$ decay can be defined
analogously to the expression for the electromagnetic form factor of
the $\pi^\pm/K^\pm$ meson, see eq. (\ref{FormFactorBSE}). In our model, it reads explicitly
\begin{eqnarray}
\left\langle\pi^-(P')\left|J_\mu^{\mbox{\tiny
weak}}\right|K^0(P)\right\rangle &&\\
= -\int\frac{d^4 p}{(2\pi)^4} \mbox{tr}&\Big[&\bar{\Gamma}_{(\pi^-)}^{P'}(p+\frac q
    2) S_d^F(\frac P 2 +
p)\Gamma _{(K^0)}^P (p) S_{\bar s}^F(-\frac P 2 + p) \bar u \gamma
_\mu s S_{\bar u}^F(-\frac P 2 + p+q)\Big] \nonumber 
\end{eqnarray} 
Note that for $0^-\to 0^-$ transitions only the vector part of the
weak coupling in $J_\mu^{\mbox{\tiny weak}}\left.\right|_{\mbox{\tiny Dirac}}=\gamma_\mu(\Id-\gamma _5)$
contributes to the form factors; furthermore, a second term with a
$W$ boson coupling to the $d$ quark trivially does not occur due to its vanishing
flavour trace.

In most experiments, the $q^2=(P'-P)^2=:-Q^2$ dependence of the $f_\pm$
form factor are found to be consistent with a linear parametrization
like
\begin{equation}
\label{LinearParametrization}
f_\pm(Q^2) = f_\pm(0)\:\left(1+\lambda_\pm\cdot Q^2\right) \qquad .
\end{equation}
Since $f_-$ is multiplied by the lepton mass, this form factor is
difficult to measure; however, there exist some rough experimental values for the ratio
$\xi:=f_-/f_+$. 

In table \ref{tab:KaonL3Decay}, we show our results for the form
factors at $Q^2=0$ and their ratio $\xi$. For $f_+$ and $f_-$, the
absolute values of these form factors are not determined so that only their ratio can be
compared with experimental estimations. 

\begin{figure}
  \protect\begin{center}
    \leavevmode
       \protect\input{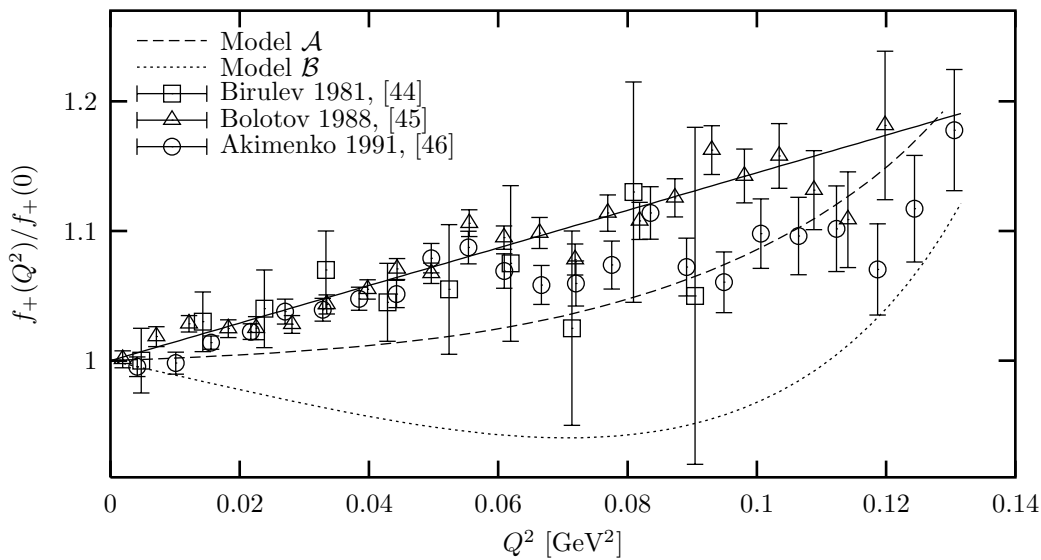}
\newline
       \protect\caption{The normalized form factor $f_+(Q^2)/f_+(0)$
  of the $K_{\ell 3}$ decay, calculated with the parameters of model
  $\cal A$ and model $\cal B$. The solid line indicates the linear fit
  according to eq. (\ref{LinearParametrization}) with
  the parameter $\lambda _+^{\mbox{\tiny PDG}}=(0.0276\pm 0.0021)/M_\pi^2$, see
  ref. \protect\cite{PDG00}.} 
      \protect \label{fig:KaonL3Decay}
   \end{center}
\end{figure}

We show the $Q^2$ dependence of the normalized $f_+$ form factor in
fig. (\ref{fig:KaonL3Decay}). Obviously, a linear fit to the
experimental data is justified only approximately; indeed, our results
show remarkable non--linear shapes. Although the data are
underestimated in model $\cal B$, they can be acceptably described with the
parameters of model $\cal A$ so that no general shortcoming of our
model can be stated. 

Let us finally note a special feature of our
model: the {\sl nonstrange} and the {\sl strange} quark sector are distinct
not only due to different constituent masses, but also because of the 't Hooft
couplings $g$ and $g'$. In the simultaneous limit $m_s\to m_n$ and
$g'\to g$, the $SU_F(3)$ symmetry is restored and thus the
kaon amplitude will become the pion amplitude. The  $SU_F(3)$
limit leads then to $-f_+(Q^2)\to F_\pi(Q^2)$ and $f_-(Q^2)\to 0$ with
$F_\pi (Q^2)$ being the well--known pion form factor, see
eq. (\ref{FormFactorDef}); this trivial result has been checked
numerically in our model.

\section{Summary and Outlook}        
                                        \label{SummaryOutlook}

In this paper, we have presented some new results of a relativistic quark
model for mesons. We have briefly resumed our approach on the basis of the
Bethe--Salpeter equation in its instantaneous approximation. The
potential in the resulting three--dimensional reduction is a combination of a
linear confinement potential plus a residual interaction \`a la 't Hooft
based on instanton effects.

Our numerical calculations were done with two different parameter sets
that have distinct Dirac structures for the confinement. The discussion of the complete
meson spectrum shows an excellent agreement with the experimental
data; the correct splittings in the pseudoscalar sector can be
backtraced to the effects of our residual interaction which in turn
yields remarkable splittings for scalar states. We found considerable
differences between the spectra of model $\cal A$ and 
model $\cal B$: the masses of all $J^\pi=0^+$ ground states were significantly
lowered in model $\cal B$ so that the assignment of
possible $q\bar q$ states in this puzzling sector will differ in both models.

Furthermore, we have investigated various meson decay modes such as
the $K_{\ell 3}$ decay and $\pi^0,\eta,\eta '\to\gamma\gamma$. The
latter transitions were studied not only for real photons but also for
very high virtualities. We found that our model fails if only one photon is
virtual: the asymptotic limit ({\it e.g.} $Q^2F^{\pi^0}_{\gamma\gamma}(Q^2,0)\to 2f_{\pi^0}$)
known from perturbative QCD is not recovered. However, we can proof
similar relations for $Q^2F^{\cal M}_{\gamma\gamma}(Q^2,Q^2)$ linking the transition form
factor of the meson ${\cal M}=\pi^0,\eta,\eta '$ to its decay
constants in the case of symmetric photon virtualities. We stress that the analytic
calculations presented here are in fact model--independent. Finally, we
found excellent agreement of the various form factors in the decays
$\pi^+/K^+\to \ell^+\nu _\ell\gamma$ as well as a satisfying
description of the electromagnetic $\pi^\pm/K^\pm$ form factors in
both parameter sets. 

The relativistic quark model presented in this paper also allows
further investigations; we have already mentioned an extension of 't
Hooft's residual interaction for heavy $q\bar q$ systems. A further
topic is the study of strong decays in this framework. Hereby, it is
of special interest that not only pure quark loops contribute to the
strong decay widths, but also instanton--induced six--quark interactions will
occur. Furthermore, we have studied the various implications of our quark
model with respect to the concept of spontaneous breaking of the
chiral symmetry; various low--energy theorems can be tested and
compared with results from Chiral Perturbation Theory. A last point to
mention is the study of Compton scattering off a pseudoscalar meson:
it is possible to calculate the corresponding matrix elements in the
framework of our Bethe--Salpeter model and compute the electromagnetic
polarizabilities, even in their generalized form for virtual
photons. All these various aspects are currently prepared for
publication and will soon be presented in subsequent papers.


\section*{Acknowledgements}

We thank E. Klempt, V. V. Anisovich and A. Sarantsev for fruitful
discussions. Financial support of the {\sc Deutsche
Forschungsgemeinschaft} (DFG) is gratefully acknowledged.


\vspace*{2cm}

\bibliographystyle{plain}


\vspace*{2cm}

\section*{Appendix}

In this Appendix, we present a model--independent factorization proof
for the asymptotic limit ($Q^2\to\infty$) of the pseudoscalar transition form factor
$Q^2F^{\cal M}_{\gamma\gamma}(Q^2,Q^2)$ at equal photon virtualities. We stress
that the following considerations are formulated on the basis of full
four--dimensional Bethe--Salpeter amplitudes and quark propagators so
that it will be independent of the instantaneous approximation that we
adopted for our numerical evaluations.

Let us start with the matrix element $T_{\gamma\gamma}^{\cal M}$ in
eq. (\ref{MatrixElement}) for the decays ${\cal M}\to\gamma\gamma$
with ${\cal M}=\pi^0,\eta,\eta '$ at arbitrary photon virtualities
$Q_i^2=-q_i^2$. We define $q:=\frac 1 2 (q_1-q_2)$ and consider the
decaying meson in its rest frame with $P=(M,\vec 0)$. With $q_1=P/2+q$
and $q_2=P/2-q$, we then find the following relations for
$Q_1^2=Q_2^2=:Q^2\to\infty$: 
\begin{equation}
\begin{array}{ccc} q^2&\to& -Q^2 \\ q^0&\to& 0\end{array}
\Longrightarrow
\begin{array}{ccc} q_1^0, q_2^0&\to& \frac M 2 \\ q_1^3, -q_2^3&\to&
\sqrt{Q^2}\end{array} 
\end{equation}
Here, we have chosen the photon momenta to be in the direction of the
$z$--axis. We now study the behaviour of the intermediate propagator
in both terms of the matrix element $T_{\gamma\gamma}^{\cal M}$ in
eq. (\ref{MatrixElement}). The
denominator behaves like $(P/2+p-q_i)^2\to -Q^2$ for asymptotic photon
virtualities since terms proportional to the relative momentum
$p$ will not contribute due to the vanishing vertex function for
$p\to\infty$. For the same reason, only a $\vec q_i \vec\gamma =
\pm\sqrt{Q^2}\gamma^3$ term survives in the numerator (``$+$'' for
$i=1$, ``$-$'' for $i=2$). Therefore we find for the complete
intermediate quark propagator
\begin{equation}
S^F(\frac P 2 + p + q_i) \to \pm\frac{\sqrt{Q^2}}{Q^2}\gamma_3
\qquad\mbox{as }\quad Q^2\to\infty\quad.
\end{equation}
Inserting this in the matrix element of eq. (\ref{MatrixElement}), we find
\begin{equation}
\label{TAsympt}
T_{\gamma\gamma}^{\cal M} \to -i\sqrt 3\:\frac{\sqrt{Q^2}}{Q^2}\int\frac{d^4
p}{(2\pi)^4}\:\mbox{tr }\left[ S^F(\frac P 2 + p)\Gamma ^P_{({\cal M})}(p)S^F(-\frac P 2 + p)\left( \not\!\varepsilon
_2 \gamma _3\not\!\varepsilon _1 - \not\!\varepsilon
_1 \gamma _3\not\!\varepsilon _2 \right)\right]
\end{equation}
for asymptotic virtualities. Since we are dealing with virtual
photons, the polarization vectors $\varepsilon _i=(0,\vec \varepsilon _i)$ do not only have
transversal but also longitudinal components. However, if one of the
photons is longitudinally polarized with $\vec\varepsilon_i\:\|\:\vec
q_i\:\|\:\vec e_3$, the two terms in eq. (\ref{TAsympt}) will cancel. The
same happens for $\vec \varepsilon _1=\vec\varepsilon _2$ so that we conclude
\begin{equation}
 \not\!\varepsilon
_2 \gamma _3\not\!\varepsilon _1 - \not\!\varepsilon
_1 \gamma _3\not\!\varepsilon _2 = \left\{
\begin{array}{cc}
0 & \qquad\mbox{if }\quad\vec\varepsilon _1=\vec\varepsilon _2\quad\mbox {or
}\quad \vec\varepsilon_i\:\|\:\vec q_i \\
\mp 2i\gamma _0\gamma _5 &\qquad \mbox{otherwise}
\end{array} \right .
\end{equation}
where, as usual, $\gamma _5=i\gamma _0 \gamma _1 \gamma _2 \gamma _3$. Here and in
the following, the upper sign (``$-$'') stands for polarization vectors $\varepsilon
_1=(0,1,0,0)$ and $\varepsilon _2=(0,0,1,0)$, while the lower sign
(``$+$'') applies to the choice $\varepsilon _1=(0,0,1,0)$ and
$\varepsilon _2=(0,1,0,0)$. 
The resulting integrand is equivalent to the expression for the pseudoscalar decay
constants in eq. (\ref{f-pi-Salpeter}) in its four--dimensional generalization for
${\cal M}=\pi^0,\eta,\eta '$. By comparison of the asymptotic matrix
element and the definition of the decay constants $f_{\cal M}^j$
($j=\pi,n,s$) we find
\begin{eqnarray} 
T_{\gamma\gamma}^{\cal M} &\to &\mp 2\sqrt 3\:\frac{\sqrt{Q^2}}{Q^2}\int\frac{d^4
p}{(2\pi)^4}\:\mbox{tr }\left[ S^F(\frac P 2 + p)\Gamma ^P_{({\cal
M})}(p)S^F(-\frac P 2 + p)\gamma _0\gamma 
_5\right] \nonumber \\ &=& \mp 2\frac{M_{\cal M}\sqrt{Q^2}}{Q^2}\sum _j\tilde C_j f_{\cal M}^j
\end{eqnarray} 
as $Q^2\to\infty$. If we consider the process $\pi^0\to\gamma\gamma$,
the sum becomes trivial 
and contains only the pion decay constant $f_\pi$ multiplied by the factor
$\tilde C_\pi = \alpha/(3\sqrt 2)$. We note that we
have introduced the charge factors $\tilde C_j:=\alpha C_j$ 
that are proportional to the factors defined in section \ref{GammaGammaDecay}. 
Now we recall the relation between the matrix element of the two
photon decay and the transition form factor in eq. (\ref{TandF}). Since we have set the
photon momenta as $q_i=(q_i^0;0,0,q_i^3)$, we can evaluate the
implicit sum over the indices $\mu,\nu,\alpha$ and $\beta$. This yields
\begin{equation} 
\mp\alpha M_{\cal M} \sqrt{Q^2}F^{\cal M}_{\gamma\gamma}(Q^2,Q^2)= \mp
2\frac{M_{\cal M}\sqrt{Q^2}}{Q^2}\sum _j\tilde C_j f_{\cal
M}^j\qquad\mbox{as }\quad Q^2\to\infty
\end{equation} 
so that we finally arrive at the asymptotic limits quoted in
eqs. (\ref{OnePoleLimitVV}) and (\ref{TwoPoleLimitVV}):
\begin{eqnarray}
\lim_{Q^2\to \infty} Q^2 F^{\pi^0}_{\gamma\gamma}(Q^2,Q^2)&=& \frac 2 3 f_{\pi^0} \\
\lim_{Q^2\to \infty} Q^2 F^{\eta}_{\gamma\gamma} (Q^2,Q^2)&=& \frac{5}{9\sqrt{2}}
f_{\eta}^n + \frac{1}{9} f_{\eta}^s \\ 
\lim_{Q^2\to \infty} Q^2 F^{\eta '}_{\gamma\gamma} (Q^2,Q^2)&=& \frac{5}{9\sqrt{2}} f_{\eta '}^n +
\frac{1}{9} f_{\eta '}^s 
\end{eqnarray}
We want to emphasize that this proof is entirely model--independent
since we only explored kinematical properties of the diagrams in
fig. (\ref{fig:TPD}) without making any assumptions on the meson
vertex function, in particular without applying the instantaneous approximation.


\newpage 

\section*{tables}

\begin{table}
  \begin{tabular}{cccc}
\protect
\rule[-6mm]{0mm}{13mm} & {\bf Parameter} & {\bf Model $\cal A$ } & {\bf
Model $\cal B$} \\
\hline 
 & & & \\
& $g$ [GeV${}^{-2}$] 		& 1.73 & 1.62  \\
\raisebox{2ex}[1ex]{\sl 't Hooft} &$g'$ [GeV${}^{-2}$] 	& 1.54 & 1.35  \\ 
\raisebox{2ex}[1ex]{\sl interaction}& $\Lambda _{\mbox{\scriptsize III}}$ [fm] & 0.30 & 0.42\\ 
 & & & \\
{\sl Constituent}& $m_n$ [MeV] & 306 & 380 \\
{\sl quark masses}&$m_s$ [MeV] & 503 & 550 \\
 & & & \\
 {\sl Confinement}& $a_c$ [MeV] & --1751 & --1135 \\
 {\sl parameters}& $b_c$ [MeV/fm] & 2076 & 1300 \\
& & & \\
{\sl Spin structure} &${\mit\Gamma\otimes\Gamma}$ & $\frac 1 2 (\Id\otimes\Id - \gamma _0\otimes\gamma _0)$ &
$\frac 1 2 (\Id\otimes\Id - \gamma _\mu\otimes\gamma ^\mu - \gamma
_5\otimes\gamma _5)$  \\
 & & & \\
  \end{tabular}
\caption{The parameters of the confinement force, the 't
Hooft interaction and the constituent quark masses in the models
$\cal A$ and $\cal B$.}
\label{tab:Parameters}
\end{table}

\begin{table}
  \begin{tabular}{ccccc|ccccc}
\protect
\rule[-6mm]{0mm}{13mm} {\bf Meson ($J^{\pi c}$)} & {\bf $n$}& {\bf Model
$\cal A$ } & {\bf Model $\cal B$} &&& {\bf Meson ($J^{\pi c}$)} & {\bf
$n$}& {\bf Model
$\cal A$ } & {\bf Model $\cal B$} \\
\hline 
& & & &&& & & & \\
                 & $0$ &   138      &    140  &&&               &  $0$    & 1240       & 1201       \\
                 & $1$ &  1357      &   1331  &&& $b_1(1^{+-})$ &  $1$    & 1876       & 1718       \\    
$\pi(0^{-+})$    & $2$ &  2012      &   1826  &&&               &  $2$    & 2373       & 2099       \\
                 & $3$ &  2498      &   2193  &&&               &           &            &            \\
                 & $4$ &  2898      &   2496  &&&               &  $0$    & 1321       & 1057       \\
                 &       &            &         &&&  $a_0(0^{++})$&  $1$    & 1931       & 1665       \\
                 & $0$ &   778      &    785  &&&               &  $2$    & 2423       & 2071       \\
                 & $1$ &  1553      &   1420  &&&               &           &            &            \\    
                 & $2$ &  1605      &   1472  &&&               &  $0$    & 1240       & 1201       \\
                 & $3$ &  2118      &   1891  &&&  $a_1(1^{++})$&  $1$    & 1876       & 1718       \\
$\rho(1^{--})$   & $4$ &  2161      &   1913  &&&               &  $2$    & 2373       & 2099       \\
                 & $5$ &  2567      &   2244  &&&               &           &            &            \\
                 & $6$ &  2608      &   2257  &&&               &  $0$    & 1312       & 1358       \\
                 & $7$ &  2949      &   2538  &&&  $a_2(2^{++})$&  $1$    & 1879       & 1768       \\
                 & $8$ &  2987      &   2547  &&&               &  $2$    & 1931       & 1807       \\
                 &                    &         &&&               &           &            &            \\
                 & $0$ &  1633      &   1605  &&&               &  $0$    & 1951       & 1926       \\
$\pi _2(2^{-+})$ & $1$ &  2156      &   1997  &&&  $a_3(3^{++})$&  $1$    & 2401       & 2247       \\    
                 & $2$ &  2592      &   2318  &&&               &  $2$    & 2792       & 2525       \\
                 &       &            &         &&&               &           &            &            \\
                 & $0$ &  1698      &   1743  &&&               &  $0$    & 2011       & 2052       \\
                 & $1$ &  2157      &   2060  &&&  $a_4(4^{++})$&  $1$    & 2402       & 2315       \\    
$\rho _3(3^{--})$& $2$ &  2208      &   2091  &&&               &  $2$    & 2451       & 2341       \\
                 & $3$ &  2576      &   2371  &&&               &           &            &            \\
                 & $4$ &  2631      &   2388  &&&               &  $0$    & 2463       & 2444       \\
                 &       &            &         &&&  $a_5(5^{++})$&  $1$    & 2825       & 2685       \\
                 & $0$ &  2279      &   2318  &&&               &  $2$    & 3153       & 2905       \\
                 & $1$ &  2623      &   2545  &&&               &           &            &            \\    
$\rho _5(5^{--})$& $2$ &  2671      &   2568  &&&               &  $0$    &  2517      & 2554       \\
                 & $3$ &  2967      &   2782  &&&  $a_6(6^{++})$&  $1$    &  2826      & 2755       \\
                 & $4$ &  3019      &   2797  &&&               &  $2$    &  2872      & 2776       \\
& & & &&& & & & \\
  \end{tabular}
\protect\caption{\protect Masses of the isovector mesons in [MeV], calculated
with the parameters of the models $\cal A$ and $\cal B$; here, $n$
denotes the radial excitation. For a
comparison with the latest experimental data of ref. \protect\cite{PDG00}, see
fig. (\protect\ref{fig:IsoVector}).}
\protect\label{tab:MassesIsoVector}
\end{table}

\begin{table}
  \begin{tabular}{ccccc|ccccc}
\protect
\rule[-6mm]{0mm}{13mm} {\bf Meson ($J^{\pi }$)} & {\bf $n$}& {\bf Model
$\cal A$ } & {\bf Model $\cal B$} &&& {\bf Meson ($J^{\pi }$)} & {\bf
$n$}& {\bf Model
$\cal A$ } & {\bf Model $\cal B$} \\
\hline 
& & & &&& & & & \\
                 & $0$ &  499      &  506  &&&               &  $0$    & 1426     &  1187     \\
                 & $1$ & 1508      & 1470  &&& $K^*_0(0^{+})$&  $1$    & 2058     &  1788     \\    
                 & $2$ & 2159      & 1965  &&&               &  $2$    & 2561     &  2196     \\
  $K(0^{-})$     & $3$ & 2652      & 2336  &&&               &           &      &       \\
                 & $4$ & 3062      & 2644  &&&               &  $0$    &  870      &   890   \\
                 & $5$ & 3420      & 2911  &&&               &  $1$    & 1687      &  1550   \\ 
                 & $6$ & 3748      & 3151  &&& $K^*(1^{-})$  &  $2$    & 1718      &  1588   \\
                 & $7$ & 4074      & 3370  &&&               &  $3$    & 2261      &  2018   \\    
                 &       &           &       &&&               &  $4$    & 2289      &  2037   \\
                 & $0$      &  1353         &  1315     &&&       &           &      &       \\
                 & $1$      &  1353         &  1315     &&&       &   $0$        & 1406     & 1447      \\
   $K_1(1^{+})$               & $2$      &  2005         &  1840     &&&  $K^*_2(2^{+})$   & $1$       & 2019     & 1889      \\
                 & $3$      &  2005         &  1840     &&&       &    $2$        & 2049     & 1914      \\
                 &       &           &       &&&       &           &      &       \\
                 & $0$        & 1750      & 1709  &&&               &    $0$       & 1800     & 1828      \\
  $K_2(2^{-})$                & $1$        & 1750      & 1709  &&&   $K^*_3(3^{-})$              & $1$        & 2300     & 2173      \\
                 & $2$        & 2292      & 2115  &&&               &    $2$        & 2334     & 2192      \\
         &       &       &   &&&               &           &      &       \\
                      & $0$  & 2074      & 2026   &&&               &   $0$        & 2121     & 2136      \\
    $K_3(3^{+})$                   & $1$  & 2074      & 2026   &&&  $K^*_4(4^{+})$              & $1$        & 2550     & 2424      \\
                      & $2$  & 2544      & 2362   &&&               &   $2$         & 2585     & 2439      \\
                 &       &       &   &&& &           &      &       \\    
                    &  $0$   & 2352      &   2299       &&&               &  $0$         & 2397     &  2400     \\
    $K_4(4^{-})$                 &  $1$   & 2352      &   2299       &&&   $K^*_5(5^{-})$              & $1$        & 2775     &  2649     \\
                    &  $2$   & 2771      &   2587       &&&               &   $2$         & 2811     &  2661     \\
& & & &&& & & & \\
  \end{tabular}
\protect\caption{\protect Masses of the isodublet mesons in [MeV], calculated
with the parameters of the models $\cal A$ and $\cal B$; here, $n$
denotes the radial excitation. For a
comparison with the latest experimental data of ref. \protect\cite{PDG00}, see
fig. (\protect\ref{fig:IsoDublett}). Note that the calculated $K_1$ states are each 2--fold degenerate for
    spin $S=0$ and $S=1$.}
\protect\label{tab:MassesIsoDublett}
\end{table}

\begin{table}
  \begin{tabular}{ccccc|ccccc}
\protect
\rule[-6mm]{0mm}{13mm} {\bf Meson ($J^{\pi c}$)} & {\bf $n$}& {\bf Model
$\cal A$ } & {\bf Model $\cal B$} &&& {\bf Meson ($J^{\pi c}$)} & {\bf
$n$}& {\bf Model
$\cal A$ } & {\bf Model $\cal B$} \\
\hline 
& & & &&& & & & \\
& $0$&   531   &  533  &&& & $0$   &  984  &   665  \\
& $1$&   975   &  950  &&& & $1$   & 1468  &  1262  \\    
& $2$&  1533   & 1446  &&& & $2$   & 1776  &  1554  \\
& $3$&  1812   & 1654  &&&$f_0(0^{++})$ & $3$   & 2113  &  1870  \\
$\eta(0^{-+})$& $4$&  2177   & 1912  &&& & $4$   & 2310  &  2006  \\
& $5$&  2381   & 2118  &&& & $5$   & 2617  &  2281  \\
& $6$&  2657   & 2267  &&& & $6$   & 2756  &  2359  \\
& $7$&  2838   & 2479  &&& &    &   &    \\    
& $8$&  3054   & 2565  &&& & $0$   & 1240  &  1201  \\
& &     &   &&&$f_1(1^{++})$ & $1$   & 1454  &  1422  \\
& $0$&   778   &  785  &&& & $2$   & 1876  &  1718  \\    
& $1$&   954   &  990  &&& &    &   &    \\
& $2$&  1553   & 1420  &&& & $0$   & 1312  &  1358  \\
& $3$&  1605   & 1472  &&& & $1$   & 1495  &  1537  \\    
$\omega/\phi(1^{--})$& $4$&  1804   & 1674  &&& & $2$   & 1879  &  1768  \\
& $5$&  1829   & 1701  &&& & $3$   & 1931  &  1807  \\
& $6$&  2118   & 1891  &&& & $4$   & 2147  &  2006  \\
& $7$&  2161   & 1913  &&& & $5$   & 2164  &  2025  \\
& &     &   &&&$f_2(2^{++})$ & $6$   & 2357  &  2141  \\
& $0$&  1633   & 1605  &&& & $7$   & 2411  &  2160  \\
$\eta _2(2^{-+})$& $1$&  1861   & 1812  &&& & $8$   & 2654  &  2392  \\
& $2$&  2156   & 1997  &&& & $9$   & 2679  &  2402  \\
& &     &   &&& & $10$  & 2761  &  2446  \\
& $0$&  1698   & 1743  &&& & $11$  & 2813  &  2457  \\    
& $1$&  1899   & 1918  &&& &    &   &    \\
$\omega _3/\phi _3(3^{--})$& $2$&  2157   & 2060  &&& & $0$   & 1951  &  1926  \\
& $3$&  2208   & 2091  &&&$f_3(3^{++})$ & $1$   & 2192  &  2127  \\
& &     &   &&& & $2$   & 2401  &  2247  \\    
& $0$& 2223    & 2200  &&& &    &   &    \\
& $1$& 2478    & 2398  &&& & $0$   & 2011  &  2052  \\
$\eta _4(4^{-+})$& $2$& 2622    & 2475  &&& & $1$   & 2230  &  2227  \\
& $3$& 2917    & 2700  &&&$f_4(4^{++})$ & $2$   & 2402  &  2315  \\
& &     &   &&& & $3$   & 2451  &  2341  \\
& $0$& 2279    & 2318  &&& &    &   &    \\    
& $1$& 2514    & 2493  &&& & $0$   & 2463  &  2444  \\
$\omega _5/\phi _5(5^{--})$& $2$& 2623    & 2545  &&&$f_5(5^{++})$ & $1$   & 2730  &  2640  \\
& $3$& 2671    & 2568  &&& & $2$   & 2825  &  2685  \\
& &     &   &&& &    &   &    \\    
& $0$& 1240    & 1201  &&& & $0$   & 2517  &  2554  \\    
$h_1(1^{+-})$& $1$& 1454    & 1422  &&&$f_6(6^{++})$ & $1$   & 2766  &  2730  \\    
& $2$& 1876    & 1718  &&& & $2$   & 2826  &  2755  \\    
& & & &&& & & & \\
  \end{tabular}
\protect\caption{\protect Masses of the isoscalar mesons in [MeV], calculated
with the parameters of the models $\cal A$ and $\cal B$; here, $n$
denotes the radial excitation. For a
comparison with the latest experimental data of ref. \protect\cite{PDG00}, see
fig. (\protect\ref{fig:IsoScalar}).}
\protect\label{tab:MassesIsoScalar}
\end{table}

\begin{table}
  \begin{tabular}{ccccccc}
\protect
\rule[-6mm]{0mm}{13mm} {\bf Coefficient} & {\bf Model $\cal A$ } & {\bf
Model $\cal B$} & {\bf Mark III , \cite{Mark3}} & {\bf DM2 ,
\cite{DM2}} & {\bf Ref. \cite{FeldmannKrollStech1}} & {\bf Ref. \cite{Cao}}\\
\hline 
 & & & \\
   $|X_{\eta }|$  & 0.68 & 0.74 & $0.67\pm 0.05$ & $0.647\pm 0.044$ & $0.774\pm 0.010$ & $0.768\pm 0.020$ \\
   $|Y_{\eta }|$  & 0.73 & 0.67 & $0.74\pm 0.10$ & $0.771\pm 0.037$ & $0.633\pm 0.013$ & $0.640\pm 0.024$ \\
 & & & \\	               	                                                                        
   $|X_{\eta '}|$ & 0.71 & 0.61 & $0.58\pm 0.06$ & $0.436\pm 0.044$ & $0.633\pm 0.013$ & $0.640\pm 0.024$ \\
   $|Y_{\eta '}|$ & 0.70 & 0.79 & $1.05\pm 0.12$ & $0.900\pm 0.021$ & $0.774\pm 0.010$ & $0.768\pm 0.020$ \\
 & & & \\
  \end{tabular}
\protect\caption{\protect The coefficients for the {\sl
nonstrange}/{\sl strange} mixing in the $\eta$ and $\eta '$ mesons according to
eq. (\ref{etaMixing}), calculated with the parameters of the models 
$\cal A$ and $\cal B$. The Mark III group used an unconstrained fit in
the evaluation of the coefficients while $X_{\cal M}^2+Y_{\cal M}^2=1$ (${\cal
M}=\eta,\etap$) was demanded in the DM2 analysis. Note that the
results of refs. \protect\cite{FeldmannKrollStech1} and
\protect\cite{Cao} are found in the so--called one--angle
mixing scheme of eq. (\protect\ref{OneAngleMixing}) where by definition $|X_{\eta }|=|Y_{\eta
'}|$ and $|Y_{\eta }|=|X_{\eta '}|$ is fixed.} 
\protect\label{tab:XYCoefficients}
\end{table}

\begin{table}
  \begin{tabular}{ccccc}
\protect
\rule[-6mm]{0mm}{13mm} {\bf Decay Constant} & {\bf Model $\cal A$ } & {\bf
Model $\cal B$} & {\bf PDG 2000, \cite{PDG00}} & {\bf Ref. \cite{FeldmannKrollStech1}}\\
\hline 
 & & & & \\
    $f_\pi$ [MeV]  	& 212   & 219    & 130.7 $\pm$ 0.46 & --- \\
 & & & & \\		                 
    $f_K$   [MeV]  	& 248   & 238    & 159.8 $\pm$ 1.88 & --- \\
 & & & & \\		                 
    $f_\eta^n$ [MeV]  	& 142   & 161    & --- &108.5 $\pm$ 2.6 \\
    $f_\eta^s$ [MeV]  	& $-205$& $-166$& --- &$-111.2$ $\pm$ 5.5 \\
 & & & & \\		                 
    $f_{\eta '}^n$ [MeV]& 92    & 95     & --- & 88.8 $\pm$ 2.5 \\
    $f_{\eta '}^s$ [MeV]& 166   & 176    & --- & 136.8 $\pm$ 6.4 \\
 & & & & \\
  \end{tabular}
\protect\caption{\protect The pseudoscalar decay constants of the $\pi$, $K$, $\eta$
and $\eta '$ mesons, calculated with the parameters of the models
$\cal A$ and $\cal B$. The results of ref. \protect\cite{FeldmannKrollStech1}
originate from a phenomenological analysis of various decay ratios,
{\it e.g.} $\Gamma(J/\psi\to\eta '\rho)/\Gamma(J/\psi\to\eta\rho)$; note
that these numerical values are found in the so--called one--angle
mixing scheme, see eq. (\ref{OneAngleMixing}).}
\protect\label{tab:DecayConstants}
\end{table}

\begin{table}
  \begin{tabular}{cccc}
\protect
\rule[-6mm]{0mm}{13mm} {\bf Decay Width} & {\bf Model $\cal A$ } & {\bf
Model $\cal B$} & {\bf PDG 2000, \cite{PDG00}} \\
\hline 
 & & & \\
    $\Gamma(\pi\to\gamma\gamma)$ [eV]   & 4.1  & 3.42 &   7.74 $\pm$ 0.56 \\
   $\Gamma(\eta\to\gamma\gamma)$ [eV]   & 215  & 213  &    460 $\pm$ 40  \\
 $\Gamma(\eta '\to\gamma\gamma)$ [eV]   & 2320 & 1480 &   4290 $\pm$ 150 \\
 & & & \\
 $\Gamma(f_0(400-1200)\to\gamma\gamma)$ [eV]   & --- & 232 &  10000$\pm$ 6000 (*)\\
 $\Gamma(f_0(980)\to\gamma\gamma)$ [eV]   & 1760 & --- &  390  $\pm$ 130   \\
 $\Gamma(a_0(980)\to\gamma\gamma)$ [eV]   & --- & 500 &  300  $\pm$ 100 (*)\\
 & & & \\
  \end{tabular}
\protect\caption{\protect The widths of the decays
$0^\pi\to\gamma\gamma$ $(\pi=\pm)$, calculated with the parameters of the models
$\cal A$ and $\cal B$. Experimental widths marked with (*) are 
quoted in ref. \protect\cite{PDG00} without using them for averages,
fits etc. The interpretation of the scalar mesons differ in models
$\cal A$ and $\cal B$ such that not all widths are calculated in
both parameter sets.} 
\protect\label{tab:GammaGammaWidths}
\end{table}

\begin{table}
  \begin{tabular}{cccc}
\protect\rule[-6mm]{0mm}{13mm}
{\bf Decay Width} & {\bf Model $\cal A$ } & {\bf Model $\cal B$} &
{\bf PDG 2000, \protect\cite{PDG00}} \\ 
\hline 
 & & & \\
 $\Gamma(\rho^\pm\to\pi^\pm\gamma)$ [keV]	   & 35.0 & 20.6       	 &  68  $\pm$ 7	 \\
 $\Gamma(\rho^0\to\pi^0\gamma	)$   [keV]	   & 35.0 & 20.6       	 &  102  $\pm$ 26 \\
 $\Gamma(\rho^0\to\eta\gamma	)$   [keV]	   & 49.7 & 39.8       	 &  36  $\pm$ 12\\
 & & & \\			                       
 $\Gamma(\omega\to\pi^0\gamma	)$ [keV]	   & 315  & 185        	 &  717  $\pm$ 42 \\
 $\Gamma(\omega\to\eta\gamma	)$ [keV]	   & 5.52 & 4.42       	 &  5.5  $\pm$ 0.8 \\
 & & & \\		
 $\Gamma(\eta '\to\rho^0\gamma	)$ [keV]	   & 87.3 & 28.0       	 &  60  $\pm$ 5	 \\
 $\Gamma(\eta '\to\omega\gamma	)$ [keV]	   & 9.70 & 3.11       	 &  6.1  $\pm$ 0.8 \\
 & & & \\		
 $\Gamma(\phi\to\eta\gamma  	)$ [keV]	   & 58.1 & 34.7       	 &  58  $\pm$ 2	 \\
 $\Gamma(\phi\to\eta '\gamma	)$ [keV]	   & 0.01 & 0.074      	 &  0.30  $\pm$ 0.16\\
 & & & \\		
 $\Gamma(K^{*\pm}\to K^\pm\gamma)$ [keV]	   & 48.0 & 28.8       	 &  50  $\pm$ 5	 \\
 $\Gamma(K^{*0}\to K^0\gamma	)$ [keV]	   & 102  & 70.2       	 & 116   $\pm$ 10 \\
 & & & \\		
 $\Gamma(b_1^\pm\to\pi^\pm\gamma)$	[keV]   & 9.21 & 7.05       	 &  230  $\pm$ 60 \\
 & & & \\		
 $\Gamma(f_1(1285)\to\rho^0\gamma)$	[keV]   & 365  & 208		& 1320   $\pm$ 310 \\
 & & & \\
  \end{tabular}
\protect\caption{\protect The widths of the decays
${\cal M}\to{\cal M}'\gamma$, calculated with the parameters of the models
$\cal A$ and $\cal B$.} 
\protect\label{tab:EMDecayWidths}
\end{table}

\begin{table}
  \begin{tabular}{ccccc}
\protect
\rule[-6mm]{0mm}{13mm} {\bf Decay Mode} & {\bf Form Factor} & {\bf Model $\cal A$ } & {\bf
Model $\cal B$} & {\bf PDG 2000, \cite{PDG00}} \\
\hline 
& & & & \\
                 & $h_V$ &0.014 &0.017 & 0.017 $\pm$ 0.008 \\
\raisebox{2ex}[1ex]{$\pi_{\ell 2\gamma}$} & $h_A$ &0.012 &0.010 & 0.0116 $\pm$ 0.0016 \\
& & & & \\
  & $h_A + h_V$   &0.124 &0.117 & 0.148 $\pm$ 0.010 \\
\raisebox{2ex}[1ex]{$K_{\ell 2\gamma}$ }  & $h_A - h_V$ &$-0.042$ &$-0.051$ & $-2.2\ldots +0.3$ \\
& & & & \\
  \end{tabular}
\protect\caption{\protect The form factors of the $\pi_{\ell 2\gamma}$
decay and the $K_{\ell 2\gamma}$ decay, calculated with the parameters
of the models $\cal A$ and $\cal B$. } 
\protect\label{tab:RadLept}
\end{table}

\begin{table}
  \begin{tabular}{crrc}
\protect
\rule[-6mm]{0mm}{13mm} {\bf Form Factor} & {\bf Model $\cal A$ } & {\bf
Model $\cal B$} & {\bf PDG 2000, \cite{PDG00}} \\
\hline 
 & & & \\
    $f_+(0)$   	& $-0.813$   	&  $-0.803$	&  --- \\
   $f_-(0)$  	& 0.121    	&  0.154   	&  ---  \\
 $\xi(0)$   	& $-0.148$ 	&  $-0.192$	&  $-0.31\pm0.15$ \\
 & & & \\
  \end{tabular}
\protect\caption{\protect The form factors of the $K_{\ell 3}$ decay
and their ratio $\xi$ at $Q^2=0$, calculated with the parameters of the
models $\cal A$ and $\cal B$.}  
\protect\label{tab:KaonL3Decay}
\end{table}


\newpage

\section*{Figures}

\protect

\protect\begin{figure}[h]
  \protect\begin{center}
    \leavevmode
       \protect\input{TwoPhotonDecay.pstex_t}
\newline
       \protect\caption{The decay ${\cal M}(P)\to\gamma _1(q_1, \epsi _1)\gamma _2(q_2,
       \epsi _2)$ in lowest order in the Mandelstam formalism.}
       \protect\label{fig:TPD}
   \end{center}
\end{figure}

\protect

\protect

\protect

\protect

\protect

\protect

\protect

\protect

\protect\begin{figure}
  \protect\begin{center}
    \leavevmode
       \protect\input{VirtualVirtualFormFactor.ModelA.latex}
\newline
       \protect\caption{The form factors of the $\gamma^*\gamma^*$
  decays at equal photon virtualities $Q^2:=Q^2_1=Q^2_2$, calculated with
  model $\cal A$, and their limits for $Q^2\to\infty$ according to the
  eqs. (\ref{OnePoleLimitVV}) and (\ref{TwoPoleLimitVV}), denoted by
  the horizontal lines.}
      \protect \label{fig:VirtualVirtualModelA}
   \end{center}
\end{figure}

\protect\begin{figure}
  \protect\begin{center}
    \leavevmode
       \protect\input{VirtualVirtualFormFactor.ModelB.latex}
\newline
       \protect\caption{The form factors of the $\gamma^*\gamma^*$
  decays at equal photon virtualities $Q^2:=Q^2_1=Q^2_2$, calculated with
  model $\cal B$, and their limits for $Q^2\to\infty$ according to the
  eqs. (\ref{OnePoleLimitVV}) and (\ref{TwoPoleLimitVV}), denoted by
  the horizontal lines.}
      \protect \label{fig:VirtualVirtualModelB}
   \end{center}
\end{figure}

\protect

\protect

\protect

\protect

\end{document}